# Analysis of the impact of heterogeneous platoon for mixed traffic flow: control strategy, fuel consumption and emissions

Yunxia Wu
School of Transportation and Logistics, National Engineering Laboratory of Integrated Transportation Big Data Application Technology, Southwest Jiaotong University, Chengdu, Sichuan 611756, China.
Email: yxwu@my.swjtu.edu.cn

Le Li
School of Transportation and Logistics, National Engineering Laboratory of Integrated Transportation Big Data Application Technology, Southwest Jiaotong University, Chengdu, Sichuan 611756, China.
Email: leli@my.swjtu.edu.cn

Zhihong Yao (Corresponding author)
School of Transportation and Logistics, National Engineering Laboratory of Integrated Transportation Big Data Application Technology, National United Engineering Laboratory of Integrated and Intelligent Transportation, Southwest Jiaotong University, Chengdu, Sichuan 611756, China.
Email: zhyao@swjtu.edu.cn

Yi Wang
School of Transportation and Logistics, National Engineering Laboratory of Integrated Transportation Big Data Application Technology, Southwest Jiaotong University, Chengdu, Sichuan 611756, China.
E-mail: wangyi1227@my.swjtu.edu.cn

Gen Li
School of Automobile and Traffic Engineering, Nanjing Forestry University, Nanjing, Jiangsu 210037, China.
Email: ligen@njfu.edu.cn

Yangsheng Jiang
School of Transportation and Logistics, National Engineering Laboratory of Integrated Transportation Big Data Application Technology, National United Engineering Laboratory of Integrated and Intelligent Transportation, Southwest Jiaotong University, Chengdu, Sichuan 611756, China.
Email: jiangyangsheng@swjtu.edu.cn

# Analysis of the impact of heterogeneous platoon for mixed traffic flow: control strategy, fuel consumption and emissions

Yunxia Wu[1,2], Le Li[1,2], Zhihong Yao[1,2,3*], Yi Wang[1,2], Gen Li[1,2], Yangsheng Jiang[1,2,3]

1. School of Transportation and Logistics, Southwest Jiaotong University, Chengdu, Sichuan 611756, China.
2. National Engineering Laboratory of Integrated Transportation Big Data Application Technology, Southwest Jiaotong University, Chengdu, Sichuan 611756, China.
3. National United Engineering Laboratory of Integrated and Intelligent Transportation, Southwest Jiaotong University, Chengdu, Sichuan 611756, China.
4. School of Automobile and Traffic Engineering, Nanjing Forestry University, Nanjing, Jiangsu 210037, China.

**Abstract**

Compared with traditional vehicle longitudinal spacing control strategies, the combination spacing strategy can integrate the advantages of different spacing control strategies. However, the impact mechanism of different combination spacing control strategies on mixed traffic flow has not been analyzed yet. Therefore, this paper proposes various combination spacing control strategies for connected automated vehicles (CAVs). First, a mixed traffic flow model was developed to analyze the characteristics of CAV platoons. On this basis, a probability model of vehicle distribution was derived, and its effectiveness was verified through simulation. Then, multiple spacing combination strategies are proposed based on four spacing control strategies. Finally, numerical experiments were conducted to calculate the average fuel consumption and pollutant emissions of mixed traffic flow under different spacing control strategies, and the impact of platoon spacing control strategies on traffic flow fuel consumption and pollutant emissions was further analyzed. Results show that: (1) the differences in average fuel consumption and pollutant emissions of traffic flow are relatively small under different platoon spacing control strategies under low traffic density (i.e., 15 veh/km); (2) at medium to high traffic densities (i.e., 55-95 veh/km), when the penetration rate of CAVs exceeds 80%, VTG1-CS, VTG2-CS, and CTG-CS strategies can effectively ensure traffic flow stability and safety, and significantly reduce fuel consumption and pollutant emissions.

Correspondence to: Zhihong Yao, School of Transportation and Logistics, Southwest Jiaotong University, Chengdu, Sichuan 610031, China, E-mail: zhyao@swjtu.edu.cn

# 1. Introduction

As an essential component of intelligent transportation systems, connected automated vehicles (CAVs) have shown broad prospects in driving safety, traffic efficiency, and emission reduction (Al-Turki et al., 2021; Chen et al., 2017; Ghiasi et al., 2017; Lioris et al., 2017; Ngoduy et al., 2021; Wang et al., 2024; Wang et al., 2021; Yao et al., 2023). Thanks to advanced sensing equipment and vehicle-to-vehicle (V2V) and vehicle-to-infrastructure (V2I) technologies, CAVs can form platoons with other CAVs to travel in smaller space, resulting in higher throughput, better stability, and lower emissions (Cai et al., 2021; Jiang, Sun, et al., 2023; Li et al., 2024; Liu et al., 2018b, 2018a; Luo et al., 2018). For CAV platoons, local stability ensures that the relative position and speed between each vehicle in the platoon remain stable (Qiang et al., 2023; Sungu et al., 2015). String stability ensures that the platoon can travel collaboratively in changing traffic environments (Guo et al., 2016; Tóth & Rödönyi, 2017; Yang et al., n.d.). Therefore, local and string stability are essential manifestations of system coordination (Bian et al., 2019; Feng et al., 2019; Swaroop & Hedrick, 1999; Zhou & Ahn, 2019). To achieve this goal, the spacing strategy is a commonly used longitudinal control method for CAVs, which adjusts the steady-state spacing between two consecutive vehicles to maintain a safe and efficient driving state. The spacing strategy has been widely studied due to its significant advantages in increasing traffic capacity (Bian et al., 2019; Yi et al., 2022; Zhang et al., 2019; Zheng et al., 2023), ensuring driving stability (Chen et al., 2019; Dong et al., 2021; Ge et al., 2024; Seiler et al., 2004; Swaroop & Hedrick, 1999; Zhang et al., 2020b), and reducing fuel consumption (Bayar et al., 2016; Mahdinia et al., 2020).

Based on different control laws, the spacing strategy can be divided into constant spacing (CS) strategy (Bian et al., 2019; Seiler et al., 2004; Swaroop & Hedrick, 1999; Zheng et al., 2016) and variable spacing strategy (Bian et al., 2019; Chen et al., 2021; Wu et al., 2020; Zhang et al., 2019; Zheng et al., 2023). Among them, variable spacing strategies include constant time gap (CTG) strategy (Bian et al., 2019; Milanés & Shladover, 2014; Zheng et al., 2023), variable time gap (VTG) strategy (Bayar et al., 2016; Chen et al., 2019, 2021; Dong et al., 2021), spacing strategy based on safety distance (SD) (Ioannou & Chien, 1993; Zhang et al., 2019), and balanced spacing (BS) strategy (Yi et al., 2022). Different spacing strategies have advantages and disadvantages (Bayar et al., 2016; Hung et al., n.d.; Zhao et al., 2009). For example, the CTG strategy can ensure string stability but requires complex communication topology. Moreover, due to the possible increase in vehicle spacing, throughput may decrease as driving speed increases (Santhanakrishnan & Rajamani, 2003). The CS-based platoon control can ensure stable high-traffic throughput but can only achieve string stability at intervals (Zheng et al., 2023). Bayar et al. (2016) show that the VTG strategy, considering the driver's driving parameters, has higher traffic capacity and lower fuel consumption than the CTG strategy. Dong et al. (2021) also compared the VTG strategy with the CTG strategy and verified that the VTG strategy can better maintain the stability of mixed traffic flow. Therefore, scholars began to explore combining different spacing strategies to utilize their respective advantages fully. Zhang et al. (2019) proposed to switch between CTG and SD strategies for vehicles at various speeds. The experimental results show that this strategy not only increases traffic flow but also improves traffic flow stability. For CAVs, Zheng et al. (2023) designed a combined spacing strategy of CTG and CS strategies. The strategy stipulates that the lead vehicle uses the CTG strategy, while the other vehicles use the CS strategy under the leader predecessor follower (LPF) communication topology. The research results also confirm that the combination of spacing strategies can effectively ensure traffic flow stability.

Numerous studies have been conducted on spacing strategies, and most scholars have focused their research on the performance of different spacing strategies in terms of stability or on the improvement of a single spacing strategy. There is still a lack of research on combined spacing strategies. Zhang et al. (2019) and Zheng et al. (2023) have studied only the combination of one spacing strategy. Moreover, this research lacks an in-depth analysis of the characteristics of mixed traffic flow. Therefore, current research has not sufficiently explored the impact of different spacing strategies on fuel consumption and emissions.

To address the gap, this paper proposes a combination of spacing strategies for CAV platoons and provides a detailed analysis of the impact of different platoon control strategies on fuel consumption and emissions in mixed traffic flow. This paper offers a theoretical basis for optimizing platoon control and traffic flow. Firstly, the characteristics of CAV platoons and the vehicle car-following patterns present in mixed traffic flow are analyzed, and the probability distribution of vehicle configuration is derived. Secondly, four spacing control strategies are introduced, and the car-following models for CAVs under different spacing strategies are analyzed. Subsequently, considering the platoon characteristics of CAVs, multiple combinations of spacing strategies for CAV platoons are proposed. Finally, the impact of different spacing combination strategies on fuel consumption and pollutant emissions is analyzed using numerical simulation. In summary, the main contributions of this paper are as follows.

(1) We propose multiple combined spacing control strategies for connected automated vehicle platoons.

(2) We analyze the composition of vehicles in mixed traffic flow and derive a probability distribution model for mixed traffic flow.

(3) We investigate the impact of different spacing control strategies on fuel consumption and pollutant emissions of mixed traffic flow under various traffic conditions.

The remainder of this paper is structured as follows. Section 2 summarizes the current research on longitudinal control strategies for CAVs. Section 3 proposes a mixed traffic flow model. Section 4 analyzes the car-following models of CAVs under different spacing control strategies and proposes multiple combined spacing strategies for CAV platoons. Section 5 examines the impact of different spacing control strategies on fuel consumption and pollutant emissions through numerical experiments. Section 6 summarizes the main findings and discusses further directions.

**2. Literature review**

Longitudinal spacing control strategies include spacing, congestion absorption, Follower-Stopper, machine learning-based control strategies, etc. Spacing is a relatively simple and easy-to-apply control strategy to set reasonable CAVs following space. The smaller the spacing setting, the higher the traffic capacity, but the risk of rear-end collision increases, and vice versa. The spacing strategy generally relies on the design of control laws to form corresponding car-following control models. It can be divided into constant spacing (CS) and variable spacing.

The CS strategy can achieve high traffic capacity as a linear control strategy. However, its requirements for vehicle tracking are rigorous, requiring more information to ensure car-following performance. Swaroop and Hedrick (1999) studied the traffic stability of the CS strategy in scenarios such as automatic control, semi-automatic control, and small platoon control. The CS strategy cannot achieve stability in the scenario of automatic control (i.e., where the vehicle can only obtain the speed and position information of the preceding vehicle through onboard sensors). The semi-automatic control scenario with the acceleration information of the preceding vehicle can ensure weak string stability. These two scenarios only consider the information of the immediately preceding vehicle, while the small platoon control scenario that finds the information of a single reference vehicle (e.g., CAVs platoon leader) has more advantages in terms of stability. Seiler et al. (2004) pointed out the need to use communication technology or nonlinear control methods to overcome the serial instability of CS strategies. From a communication perspective, generating an LPF communication topology by broadcasting information about the leading vehicle is a relatively good solution (Bian et al., 2019). Zheng et al. (2016) analyzed and compared platoon control methods with different topological structures under fixed spacing. Zhang et al. (2020a) found that the stability of the LPF platoon is sensitive to communication delay and sensor delay, so they proposed a CS strategy that considers delay. This strategy also has good string stability. Overall, the CS strategy cannot adapt to complex driving environments alone, such as frequent acceleration and deceleration of the preceding vehicle. Therefore, it is usually used in combination with other control strategies.

The variable spacing strategy is an improvement of the CS strategy, including the constant time gap (CTG) strategy, variable time gap (VTG) strategy, safety distance (SD) based spacing strategy, human driving behaviour-based spacing strategy, and so on (Wu et al., 2020; Zhang et al., 2019). It is the focus of research and application of spacing strategy. The CTG strategy is a commonly used variable spacing strategy adopted by most commercially deployed CAVs. The adaptive cruise control (ACC) and cooperative adaptive cruise control (CACC) models proposed by the PATH laboratory at the University of California also adopt the CTG strategy (Milanés & Shladover, 2014), and these two car-following control models are highly sought-after in the research field. Bian et al. (2019) extended the CTG strategy under the preceding car-following topology to a multi-preceding car-following topology. They proved that increasing the number of preceding vehicles in vehicle control can reduce the spacing between vehicles, thereby improving road transportation capacity. Zheng et al. (2023) designed a CAV platoon combination spacing strategy. This strategy assumes that the leading vehicle uses the CTG strategy and the others use the CS strategy. This combination spacing strategy can achieve high traffic capacity while ensuring system stability.

The VTG strategy mainly determines the required following spacing or time gaps through a nonlinear function of speed, which is more complex than the CTG strategy and poses a challenge to the rigorous stability analysis of the CAVs platoon. However, it can adapt flexibly to complex driving conditions, so it has also attracted the attention of many scholars. Chen et al. (2021) embedded the VTG strategy into consistency algorithms for CAV platoon control. Bayar et al. (2016) found that the VTG strategy, considering driver driving parameters, has better traffic capacity and lower fuel consumption than the CTG strategy. Dong et al. (2021) pointed out that the ACC system mainly adopts the CTG and VTG strategies, and the VTG strategy is superior to the CTG strategy in stabilizing mixed traffic flow. Chen et al. (2019) also compared the VTG and CTG strategies and verified that the VTG strategy perform better. In addition, by analyzing the braking process of vehicles in emergencies, Ioannou and Chien (1993) designed a spacing strategy based on safe distance. Zhang et al. (2019) combined the CTG and SD strategies, allowing vehicles to switch between the two strategies at different speeds. The simulation experiment results show that this combination spacing strategy increases traffic flow and enhances traffic flow stability. The ACC system needs to operate like human driving behavior to improve passenger comfort and acceptance. Based on this, Fancher et al. (2003) used the Intelligent cruise control field operational test (ICCFOT) database from the University of Michigan Transportation Research Institute to fit a distance strategy based on human driver behaviour using a quadratic curve. This study recorded the driving behaviour of 107 drivers, collected near-steady state data, and determined the pattern of human drivers controlling the distance between vehicles.

The above studies all utilize the motion information of the vehicles ahead. Other studies have found that bidirectional following communication topologies in connected environments have a balancing effect in improving stability and system robustness. Yi et al. (2022) improved the Intelligent Driver Model (IDM) by considering the difference in following distance between front and rear drivers, and proposed a bidirectional distance-balanced following model (BDBM) suitable for CAVs. This model can improve traffic flow stability while ensuring traffic efficiency. This paper names the spacing strategy that considers rear vehicle information as a balanced spacing (BS) strategy. The basic idea of this strategy is to make the vehicle travel as far as possible in the "middle" between the front and rear vehicles, achieving a balanced distance between the front and rear vehicles. Due to the additional consideration of rear vehicle information in the BS strategy, the complexity of deriving stability conditions has dramatically increased, resulting in relatively less theoretical research.

Through the above literature review, it can be found that CS strategy, CTG strategy, VTG strategy, and BS strategy have apparent advantages and disadvantages. Therefore, this paper proposes multiple combination spacing strategies to combine the benefits of different spacing strategies. Then, their comprehensive performance in traffic operation efficiency, traffic flow stability, safety, energy consumption, and pollutant emissions are investigated.

**3. Mixed traffic flow model**

3.1. Assumptions

**Assumption 1:** A connected environment without communication delay. Human-driven vehicles are also equipped with communication devices, and CAVs can obtain their acceleration information through V2V communication. However, the driving behaviour of human drivers is not influenced by communication information (Mahmassani, 2016; Yao et al., 2023) and is still subjectively determined by it. Moreover, communication delay is not considered in the CAVs control model. This means we assume a communication delay of 0 seconds.

**Assumption 2:** Vehicles drive on a single lane on a highway segment, with only longitudinal car-following considered.

**Assumption 3:** In the initial state, adjacent CAVs have been connected to form a flexible platoon.

**Assumption 4:** CAVs have high-precision sensors that can constantly sense the relative position and speed of the preceding vehicle.

**Assumption 5:** The expected acceleration output by the CAV upper controller can be instantly transmitted to the lower actuator.

**Assumption 6:** All CAVs have achieved Level 5 fully autonomous driving.

3.2. Connected automated vehicle platoon

CAVs can share real-time motion information with other vehicles through the onboard network, thereby connecting to form a flexible platoon. This strategy of driving in platoons has been proven to increase the controllability and organization of the transportation system, improve traffic safety, and achieve global optimization (Axelsson, 2017; Hall & Chin, 2005). Existing research has defined two characteristic indicators for the CAV platoon: one is platoon size $S$ and the other is platoon intensity $O$.

3.2.1 Platoon size

Due to communication limitations, the number of vehicles in the CAV platoon cannot be too large. Still, the insufficient number of vehicles also limits the advantages of the CAV platoon. Existing studies have shown that traffic capacity increases with platoon size (Chen et al., 2017; Zhou & Zhu, 2021). However, excessive platoon size negatively impacts traffic flow stability (Sala & Soriguera, 2021; Zhu & Tasic, 2021). To solve this problem, Zhu and Tasic (2021) discussed the effect of maximum platoon size on the capacity and stability of mixed traffic flow. The results showed that when the platoon size is 4, it can enhance the traffic capacity while maintaining the stability of the system. Therefore, this paper takes the maximum platoon size as 4.

3.2.2 Platoon intensity

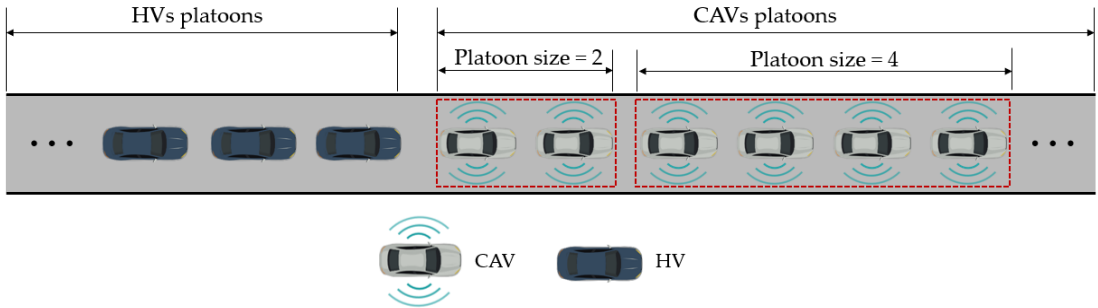

Fig. 1. The distribution of mixed traffic flow ($O = 1$).

The platoon intensity describes the degree of spatial clustering of CAVs. This concept was first proposed by Ghiasi et al. (2017) and assigned three extreme values: -1, 0, and 1. When $O = -1$, all CAVs on the road are traveling alone, and there is no situation where the size of platoons is greater than or equal to 2; when $O = 1$, all CAVs form the same platoon; when $0 = 0$, CAVs are randomly distributed on the road segment. However, when the mixed traffic flow state does not belong to the above three situations, the corresponding platoon intensity cannot be calculated. To address this issue, Jiang et al. (2023) redefined a more explanatory platoon intensity, represented by the ratio of CAVs comprising a platoon to the total number of CAVs in mixed traffic flow. The range of values for platoon intensity under this definition is from 0 to 1. This paper compares the effectiveness of different control strategies in improving mixed traffic flow. Therefore, the maximum platoon intensity value is 1, meaning all CAVs form adjacent platoons, as shown in Fig. 1.

3.3. Vehicle composition in mixed traffic flow

As shown in Fig. 2, this paper categorizes vehicles under the mixed traffic flow into four types: HV, LV1, LV2, and PV. Among them, HV vehicles are human-driven vehicles driven by human drivers. The last three types are all CAVs, driven by the auto-drive system according to the set procedures and relevant commands. LV1 vehicles are the leading vehicles of the platoon following HV vehicles, while LV2 vehicles generated due to platoon size limitations are the leading vehicles of the platoon following CAVs. Therefore, LV1 and LV2 are collectively referred to as LV vehicles. PV vehicles are the following vehicles within the platoon.

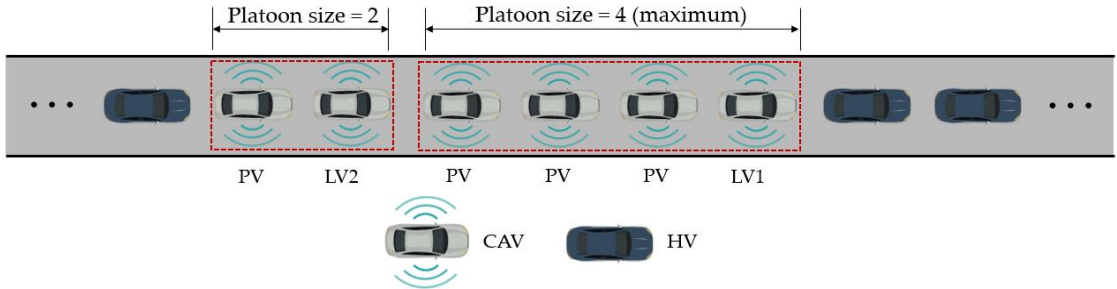

Fig. 2. The vehicle composition of mixed traffic flow.

Based on the above vehicle composition, it can be inferred that two main car-following patterns exist in mixed traffic flow.

(1) HV car-following mode

The car-following mode includes four situations: HV following HV, HV following LV1, HV following LV2, and HV following PV. HV vehicles also have communication functions in this mode but can exchange information with surrounding vehicles. However, according to Assumption 1, this information has no impact on the driving behaviour of human drivers. They still make driving decisions based on intuitive judgments of the surrounding environment. Because the actual vehicle control is in the hands of human drivers, HV vehicles cannot form synchronized driving with other vehicles.

(2) CAV car-following mode

The car-following mode includes five situations: LV1 following HV, PV following LV1, PV following LV2, LV2 following PV, and PV following PV. Due to the communication capability between the front and rear vehicles, information exchange and sharing are achieved based on V2V technology. Therefore, CAVs can obtain real-time information on the position, speed, and acceleration of surrounding vehicles, enabling synchronous driving.

3.4. Probability distribution model

In mixed traffic flow, given a CAV penetration rate of $p$, the probability of HVs is $1 - p$. To consider the platoon size and intensity of CAVs, Jiang et al. (2023) derived the distribution probabilities of five types of vehicles under mixed traffic flow based on Markov chain theory. Based on the composition of mixed traffic flow and referring to the distribution probabilities of vehicle configurations (Jiang, Zhu, Gu, et al., 2023), the distribution probabilities of different vehicles can be obtained. Denoted by $P_{LV1}$, $P_{LV2}$, and $P_{PV}$ for the distribution probabilities of LV1, LV2, and PV vehicles in the platoon, respectively,

$$P_{LV1} = (1 - p)t_{HA}, \quad (1)$$

$$P_{LV2} = \begin{cases} \dfrac{p}{S}, & O = 1 \\ \dfrac{t_{AA}^S(1 - p)t_{HA}}{1 - t_{AA}^S}, & 0 \leq O < 1 \end{cases}, \quad (2)$$

$$P_{PV} = \begin{cases} \dfrac{(S - 1)p}{S}, & O = 1 \\ \dfrac{t_{AA}(1 - t_{AA}^{S-1})(1 - p)t_{HA}}{t_{AH}(1 - t_{AA}^S)}, & 0 \leq O < 1 \end{cases}, \quad (3)$$

where $t_{UV}$ is the probability that the following vehicle is a $V$ type when the current vehicle is a $U$ type. Among them, the HV type is represented by $H$, while $A$ represents CAVs. Correspondingly, there are four car-following modes, whose probability expressions are

$$\begin{cases} t_{AH} = (1 - O)(1 - p) \\ t_{AA} = 1 - t_{AH} \\ t_{HA} = (1 - O)p \\ t_{HH} = 1 - t_{HA} \end{cases}. \quad (4)$$

When the platoon intensity value is 1, the distribution probabilities of LV1, LV2, and PV vehicles are

$$\begin{cases} P_{LV1} = 0 \\ P_{LV2} = \dfrac{p}{S} \\ P_{PV} = \dfrac{(S - 1)p}{S} \end{cases}. \quad (5)$$

The distribution probability of the LV in the platoon is

$$P_{LV} = P_{LV1} + P_{LV2} = \dfrac{p}{S}. \quad (6)$$

## 3.5. Simulation verification

To verify the effectiveness of the vehicle distribution probability derived in Section 3.4, a single-lane highway was simulated with a total of 100 vehicles and a platoon size of 4. The penetration rate increased from 0.01 to 0.99 with a step size of 0.01. The theoretical and simulation curves of vehicle distribution probabilities under different penetration rates were plotted for platoon intensity of 0 and 1, respectively, as shown in Fig. 3 and Fig. 4.

Fig. 3 shows that when the platoon intensity is 0, all vehicles are randomly distributed in the road segment. In this case, the probability of LV1 increases first and then decreases with the increase in the penetration rate of CAVs. When the penetration rate reaches about 0.5, the probability of LV1 appearing peaks. Fig. 3 (b) displays that the probability of LV2 appearing before the penetration rate is less than 0.5 is almost zero. As the penetration rate of CAVs increases from 0.5 to 0.99, the probability of LV2 distribution gradually increases, but the highest will not exceed 0.3. As shown in Fig. 3 (c), the probability of PV significantly increases with the penetration rate of CAVs. When the penetration rate approaches 0.99, the probability of PV appearing approaches 0.8. When the platoon intensity is 1, all CAVs and HVs form a platoon to travel on the road segment. In this case, the probability of LV1 appearing is almost 0. The theoretical distribution probability curves of LV2 and PV vehicles are roughly the same as the trend when the platoon intensity is 0.

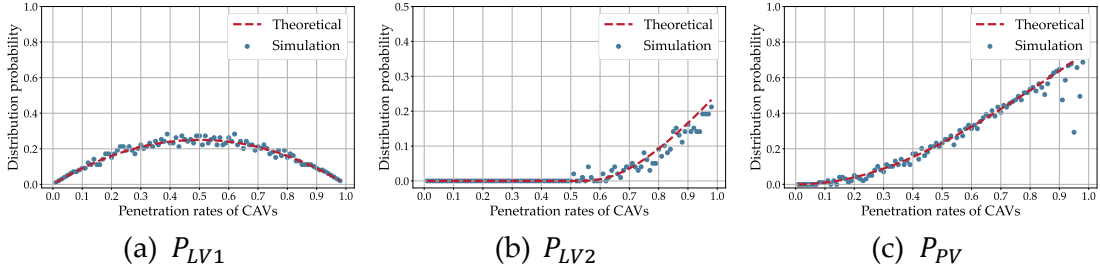

(a) $P_{LV1}$      (b) $P_{LV2}$      (c) $P_{PV}$

Fig. 3. Comparison of simulation and theoretical results ($O = 0$).

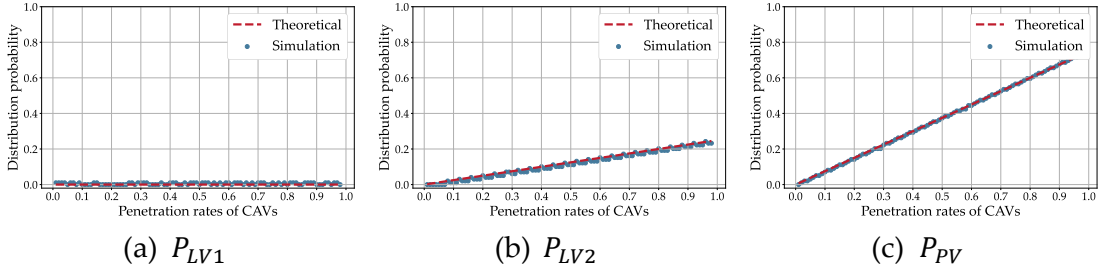

(a) $P_{LV1}$      (b) $P_{LV2}$      (c) $P_{PV}$

Fig. 4. Comparison of simulation and theoretical results ($O = 1$).

In order to verify the accuracy of the probability distribution model, the $R^2$ values of the simulation results and theoretical results are listed in Table 1. It can be seen that all $R^2$ are close to 1. This indicates that the difference between theoretical and simulated values is minimal, further verifying the correctness and effectiveness of the probability distribution model in Section 3.4. It should be noted that due to the theoretical value of the distribution probability of LV1 being 0 when the platoon intensity is 0, the simulation results are concentrated around 0. This data feature leads to the value of $R^2$ being 0. Therefore, the root mean square error (RMSE) is used in Table 1 to evaluate the difference between the simulation results and the theoretical values in this scenario.

Table 1. The performance of fitting.

| Platoon intensity | $P_{LV1}$ | $P_{LV2}$ | $P_{PV}$ |
|---|---|---|---|
| $O = 0$ | 0.9495 | 0.9423 | 0.9448 |
| $O = 1$ | 0.007* | 0.9715 | 0.9995 |

* means root mean square error (RMSE).

## 4. Vehicle car-following (control) models

4.1. Constant spacing

The CS strategy stipulates that the expected following spacing between adjacent vehicles is a fixed value; that is, the expected spacing between the target vehicle $i$ and the preceding vehicle $i-1$, denoted by $D^*_{i,CS}(t)$. It can be obtained by

$$D^*_{i,CS}(t) = d_{i,i-1} + d_0, \qquad (7)$$

where $d_{i,i-1}$ is a parameter of the CS strategy, with a value of 0 m in this paper; $d_0$ represents the minimum safe spacing.

The spacing error $e_{i,CS}(t)$ between the target vehicle $i$ and the preceding vehicle $i-1$ is calculated.

$$e_{i,CS}(t) = D_i(t) - D^*_{i,CS}(t) = x_{i-1}(t) - x_i(t) - L - (d_{i,i-1} + d_0), \qquad (8)$$

where $D_i(t)$ represents the distance between vehicle $i$ and preceding vehicle $i-1$ at time $t$; $x_{i-1}(t)$ and $x_i(t)$ represent the positions of vehicle $i-1$ and $i$, respectively; $L$ represents the length of the vehicle.

Table 2. The controller parameter of CS strategy

| Parameter | Value |
|---|---|
| $q_1$ | 0.4 |
| $q_2$ | 0.1 |
| $q_3$ | 0.9 |
| $q_4$ | 0.6 |

Under the LPF communication topology, considering the control law of the leading vehicle information in the platoon can ensure that the CS strategy has more advantages in terms of stability. Based on this control law, the expected acceleration of the target vehicle $i$ under the CS strategy $u_{i,CS}(t)$ is obtained.

$$\begin{aligned}
u_{i,CS}(t) &= \frac{1}{1+q_3}\left[a_{i-1}(t) + q_3 a_{leader}(t) + (q_1+q_2)e_{i,CS}(t)\right] \\
&\quad + \frac{1}{1+q_3}\left[q_1 q_2 e_{i,CS}(t) + (q_4+q_2 q_3)e_{i,leader,CS}(t)\right. \\
&\quad \left. + q_2 q_4 e_{i,leader,CS}(t)\right] \\
&= \frac{1}{1+q_3}\left(a_{i-1}(t) + q_3 a_{leader}(t)\right) \\
&\quad + \frac{1}{1+q_3}\left[(q_1+q_2)(v_{i-1}(t) - v_i(t))\right] \\
&\quad + \frac{1}{1+q_3}\left[q_2 q_1 (x_{i-1}(t) - x_i(t) - L - d_{i,i-1} - d_0)\right] \\
&\quad + \frac{1}{1+q_3}\left[(q_4+q_2 q_3)(v_{leader}(t) - v_i(t))\right] \\
&\quad + \frac{q_2 q_4}{1+q_3}\left[x_{leader}(t) - x_i(t) - \sum_{j=leader+1}^{i}(L + d_{i,i-1} + d_0)\right],
\end{aligned} \tag{9}$$

where $q_1$, $q_2$, $q_3$, and $q_4$ are controller parameters, their values are shown in Table 2 (Zheng et al., 2023); $leader$ is the lead vehicle code of the platoon where vehicle $i$ is located; $a_{i-1}(t)$ represents the acceleration of vehicle $i-1$ at time $t$; $v_{leader}(t)$ represents the speed of the leader vehicle at time $t$; $v_i(t)$ represents the speed of vehicle $i$ at time $t$; $x_{leader}(t)$ represents the position of the leader vehicle at time $t$.

4.2. Constant time gap

The CTG strategy requires a constant expected time gap between vehicles, so that the expected spacing between vehicles follows the speed linearly, i.e., the expected spacing $D^*_{i,CTG}(t)$ between the target vehicle $i$ and the preceding vehicle $i-1$.

$$D^*_{i,CTG}(t) = v_i(t)h + d_0, \tag{10}$$

where $h$ is a constant time gap, $h = 1.1\,s$ when the vehicle is the leading vehicle in the platoon, and $h = 0.6\,s$ when it is the following vehicle in the platoon, that is, $h_{LV} = 1.1$ s and $h_{PV} = 0.6$ s.

The spacing error $e_{i,CTG}(t)$ between the target vehicle $i$ and the preceding vehicle $i-1$ is

$$e_{i,CTG}(t) = D_i(t) - D^*_{i,CTG}(t) = x_{i-1}(t) - x_i(t) - L - v_i(t)h - d_0. \tag{11}$$

Based on the acceleration feedback control of the immediately preceding vehicle, a CAV car-following model using the CTG strategy is obtained.

$$\begin{aligned}
u_{i,CTG}(t) &= k_{e,CTG}e_{i,CTG}(t) + k_{v,CTG}(v_{i-1}(t) - v_i(t)) + k_{CTG}a_{i-1}(t) \\
&= k_{e,CTG}(x_{i-1}(t) - x_i(t) - L - v_i(t)h - d_0) \\
&\quad + k_{v,CTG}(v_{i-1}(t) - v_i(t)) + k_{CTG}a_{i-1}(t).
\end{aligned} \tag{12}$$

where $k_{e,CTG}$, $k_{v,CTG}$, and $k_{CTG}$ represent the control parameters of the inter-vehicle distance term, the control parameters of the speed difference term, and the acceleration feedback parameters under the CTG strategy, respectively.

Some works have explored the parameters of CAV controllers under the CTG strategy (Qin, 2019). Based on Qin (2019), this paper determines the values of controller parameters under the CTG strategy as follows: $k_{e,CTG} = 0.1$, $k_{v,CTG} = 0.98$, and $k_{CTG} = 0.7$.

4.3. Variable time gap

As an improvement of the CTG strategy, the VTG strategy can flexibly adjust the following time or spacing gap according to actual traffic conditions and is a nonlinear control strategy. According to existing research (Bayar et al., 2016; Chen et al., 2019, 2021; Dong et al., 2021; Yang et al., 2020), there are many forms of VTG strategies, such as VTG considering the speed of the leading vehicle in the platoon under LPF topology, VTG of segmented function type, VTG considering the information of the immediately preceding vehicle (e.g., speed of the preceding vehicle), VTG considering only the speed of the own vehicle, and so on. This paper mainly selected two of them for research.

4.3.1 The VTG strategy with the relative speed of the preceding vehicle (VTG1)

Considering the relative speed of the preceding vehicle, the VTG strategy (Yang et al., 2020) is developed. In this paper, the VTG strategy with the relative speed of the preceding vehicle is namely VTG1.

The expected spacing $D^*_{i,VTG1}(t)$ between the target vehicle $i$ and the preceding vehicle $i-1$ under the VTG1 strategy is

$$D^*_{i,VTG1}(t) = v_i(t)h_i(t) + d_0, \tag{13}$$

$$h_i(t) = c_1 - \mu\left(\frac{v_{i-1}(t)}{v_i(t)} - 1\right), \tag{14}$$

$$c_1 > 2\eta - \min\left(\mu, \frac{L + d_0}{v_f}\right), \tag{15}$$

where $v_i(t)$ and $v_{i-1}(t)$ represent the speed of vehicle $i$ and vehicle $i-1$ at time $t$, respectively; $h_i(t)$ is the expected time gap of vehicle $i$ at time $t$; $c_1$ and $\mu$ are the parameters of the VTG strategy, with $c_1 = 0.6$ s and $\mu = 0.1$ s; $\eta$ represents the engine constant, with $\eta = 0.3$ s; $v_f$ represents the free-flow speed.

Based on the above VTG strategy, the spacing error $e_{i,VTG1}(t)$ between the target vehicle $i$ and the preceding vehicle $i-1$ is derived.

$$\begin{aligned} e_{i,VTG1}(t) &= D_i(t) - D^*_{i,VTG1}(t) \\ &= x_{i-1}(t) - x_i(t) - L - (c_1 + \mu)v_i(t) + \mu v_{i-1}(t) - d_0. \end{aligned} \tag{16}$$

According to the control law, the expected acceleration of target vehicle $i$ under the VTG1 strategy is

$$\begin{aligned} u_{i,VTG1}(t) &= k_{e,VTG1}e_{i,VTG1}(t) + k_{v,VTG1}\big(v_{i-1}(t) - v_i(t)\big) + k_{VTG1}a_{i-1}(t) \\ &= k_{e,VTG1}[x_{i-1}(t) - x_i(t) - L - (c_1 + \mu)v_i(t) + \mu v_{i-1}(t) - d_0] \\ &\quad + k_{v,VTG1}\big(v_{i-1}(t) - v_i(t)\big) + k_{VTG1}a_{i-1}(t). \end{aligned} \quad (17)$$

where $k_{e,VTG1}$, $k_{v,VTG1}$, and $k_{VTG1}$ represent the control parameters of the inter-vehicle distance term, the control parameters of the speed difference term, and the acceleration feedback parameters under the VTG1 strategy, respectively.

It can be seen that the car-following model under the VTG strategy is still a function of the target vehicle's speed, the spacing between the front and rear of the vehicle, the relative speed with the front vehicle, and the acceleration of the front vehicle.

4.3.2 The VTG strategy with vehicle speed (VTG2)

The VTG strategy proposed by Yang et al. (2017) does not require consideration of the speed information of the preceding vehicle. In this strategy, the time gap is a nonlinear function related to the speed of the vehicle. In this paper, the VTG strategy with vehicle speed is namely VTG2.

$$h_i(t) = \frac{d_{VTG2}}{v_i(t)} \exp\left(\frac{v_i(t)}{2m}\right). \quad (18)$$

where $d_{VTG2}$ is the spacing parameter of the VTG strategy, $d_{VTG2} = d_0 + L = 7$ m; $m$ is the speed parameter of the VTG strategy, $m = 8.83$ m/s.

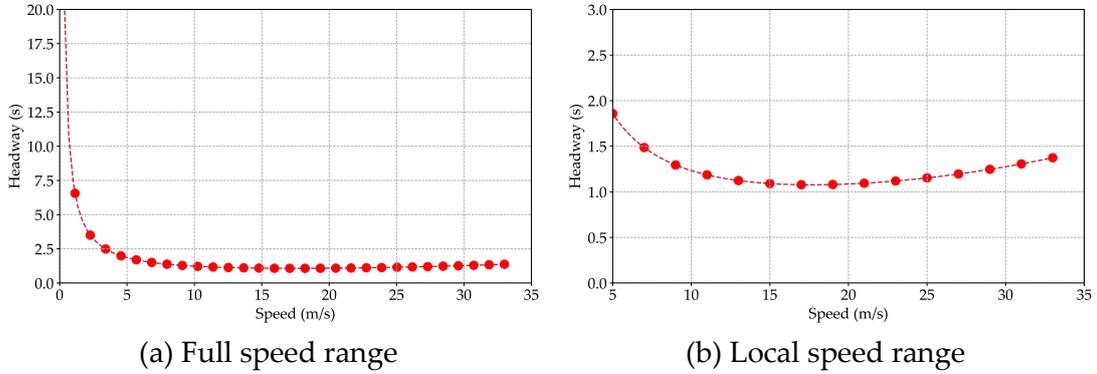

(a) Full speed range          (b) Local speed range
Fig. 5. The time headway with different speeds.

Fig. 5 shows the speed v.s. time headway. It can be seen that under this strategy, the time headway shows a trend of first sharply decreasing and then slowly increasing with the increase in speed. When the speed approaches 0, the time headway tends to infinity. This indicates that the vehicle is approaching a stationary stop. When the vehicle is driving at high speed, the required time headway is more significant than at medium speed to widen the spacing between vehicles and improve traffic safety.

The expected following spacing of the target vehicle $i$ is derived.

$$D^*_{i,VTG2}(t) = d_{VTG2} \exp\left(\frac{v_i(t)}{2m}\right) - L. \quad (19)$$

Correspondingly, the spacing error $e_{i,VTG2}(t)$ between the target vehicle $i$ and the preceding vehicle $i-1$ is

$$e_{i,VTG2}(t) = D_i(t) - D^*_{i,VTG2}(t) = x_{i-1}(t) - x_i(t) - L - \left(d_{VTG2} \exp\left(\frac{v_i(t)}{2m}\right) - L\right) \\ = x_{i-1}(t) - x_i(t) - d_{VTG2} \exp\left(\frac{v_i(t)}{2m}\right). \quad (20)$$

According to the control law, the car-following model formula for the VTG strategy is obtained.

$$u_{i,VTG2}(t) = k_{e,VTG2} e_{i,VTG2}(t) + k_{v,VTG2}(v_{i-1}(t) - v_i(t)) + k_{VTG2} a_{i-1}(t) \\ = k_{e,VTG2}\left(x_{i-1}(t) - x_i(t) - d_{VTG2} \exp\left(\frac{v_i(t)}{2m}\right)\right) \\ + k_{v,VTG2}(v_{i-1}(t) - v_i(t)) + k_{VTG2} a_{i-1}(t). \quad (21)$$

where $k_{e,VTG2}$, $k_{v,VTG2}$, and $k_{VTG2}$ represent the control parameters of the inter-vehicle distance term, the control parameters of the speed difference term, and the acceleration feedback parameters under the VTG2 strategy, respectively.

When considering time delay (i.e., communication delay and sensor delay) in the car-following model, this VTG2 strategy that only finds the speed of the vehicle can significantly simplify the complexity of stability analysis. This paper does not consider lateral control, but the analysis of the VTG2 strategy can lay the foundation for subsequent research.

4.3.3 The controller parameters

To better compare the advantages and disadvantages of various spacing strategies, this paper uses control laws that consider the acceleration feedback of the adjacent preceding vehicle. Referring to the parameter values related to the CTG strategy (Qin, 2019), this paper maximizes and unifies the controller parameter values under the VTG strategy, as shown in Table 3.

Table 3. The controller parameters of VTG.

| VTG | $k_e$ | $k_v$ | $k$ |
|---|---|---|---|
| VTG1 | 0.1 | 0.98 | 0.7 |
| VTG2 | 0.1 | 0.98 | 0.7 |

In order to ensure the rationality of the parameters, stability tests need to be conducted on the values of each parameter. For the CAV homogeneous traffic flow, the speed disturbance transfer function $G_i(s)$ when the speed disturbance experienced by vehicle $i$ is

$$G_i(s) = \frac{\tilde{V}_i(s)}{\tilde{V}_{i-1}(s)} = \frac{k_i s^2 + g_i^{\Delta v} s + g_i^{\Delta x}}{s^2 + (g_i^{\Delta v} - g_i^v)s + g_i^{\Delta x}}, \quad (22)$$

where $\tilde{V}_i(s)$ and $\tilde{V}_{i-1}(s)$ are the Laplace transforms of the speed disturbances of vehicle $i$ and $i-1$, respectively; $k_i$ represents the feedback control coefficients of the acceleration of the preceding vehicle; $s$ is the Laplace domain. Let $s = j\omega(\omega \geq 0)$ to transform the transfer function from the Laplace domain to the frequency domain. $g_i^v$ is the general expression of the car-following model being the partial derivative of the equilibrium state for the speed of the vehicle; $g_i^{\Delta x}$ is the general expression of the car-following model being the partial derivative of the spacing in the equilibrium state; $g_i^{\Delta v}$ is the general expression of the car-following model being the partial derivative of relative velocity in the equilibrium state. The relevant calculation equations for the three partial derivatives mentioned above are

$$\begin{cases} g_i^v = \left.\dfrac{\partial g(v_i, \Delta x_i, \Delta v_i)}{\partial v_i}\right|_{(v_e, \Delta x_e, 0)} \\ g_i^{\Delta x} = \left.\dfrac{\partial g(v_i, \Delta x_i, \Delta v_i)}{\partial V x_i}\right|_{(v_e, \Delta x_e, 0)} \\ g_i^{\Delta v} = \left.\dfrac{\partial g(v_i, \Delta x_i, \Delta v_i)}{\partial V v_i}\right|_{(v_e, \Delta x_e, 0)} \end{cases}. \tag{23}$$

According to the principles of control theory, if the maximum amplitude of the speed disturbance transfer function in the frequency domain does not exceed 1, the speed disturbance will not be amplified in the vehicle transmission, and the traffic flow will be in a stable state. Therefore, the criteria for determining the stability of CAV homogeneous traffic flow is

$$\|G(j\omega)\|_\infty = \left\|\prod_{i=1}^{N} G_i(j\omega)\right\|_\infty \leq 1, \tag{24}$$

where $\|\cdot\|_\infty$ is the maximum amplitude of the transfer function in the frequency domain; $N$ is the number of vehicles in the traffic flow.

When all CAVs in the traffic flow adopt the same control strategy, that is, when the control parameters of the car-following model are entirely consistent, the above stability discrimination criteria can be simplified as

$$\|G_{i,policy}(j\omega)\|_\infty = \left\|\frac{\left(g_{i,policy}^{\Delta x} - k_{i,policy}\omega^2\right) + j\omega g_{i,policy}^{\Delta v}}{\left(g_{i,policy}^{\Delta x} - \omega^2\right) + j\omega\left(g_{i,policy}^{\Delta v} - g_{i,policy}^v\right)}\right\|_\infty \leq 1. \tag{25}$$

where $g_{i,policy}^{\Delta x}$ and $k_{i,policy}$ represent the partial derivatives of certain policy for spacing in equilibrium and the feedback control coefficients of the acceleration of the preceding vehicle, respectively; $policy = \{CS, CTG, VTG1, VTG2\}$.

By solving the above inequality, we can obtain

$$\begin{cases} 0 \leq k_{i,policy} \leq 1 \\ \left(g_{i,policy}^v\right)^2 - 2g_{i,policy}^v g_{i,policy}^{\Delta v} - 2(1 - k_{i,policy})g_{i,policy}^{\Delta x} \geq 0 \end{cases}. \tag{26}$$

According to the derived CAV homogeneous traffic flow stability condition Eq. (26), the parameter values of CAV controllers under VTG strategies were tested, and the calculation results are shown in Table 4. It can be seen that the values of the controller parameters in Table 4 can ensure the stability of the vehicle, so the values are reasonable.

Table 4. Test results of CAV control parameter values.

| Spacing strategy | $k$ in [0,1] | Stability condition | Stability |
|---|---|---|---|
| VTG1 | ✓ | 0.0624 | ✓ |
| VTG2 | ✓ | Fig. 6 | ✓ |

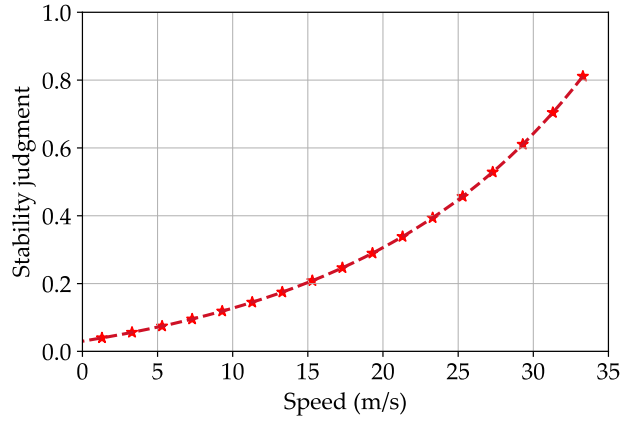

Fig. 6. The Stability condition of VTG2.

### 4.4. Balanced spacing

The balanced spacing (BS) strategy emphasizes that vehicles should travel as close as possible between the preceding and following vehicles. As a result, under this strategy, vehicles not only need to obtain the motion information of the preceding vehicle, but also need additional information from the following vehicle. Yi et al. (2022) proposed a bidirectional distance-balanced car-following model (BDBM) based on the IDM,

$$S_i(t) = d_0 + v_i(t)T - \frac{v_i(t)\Delta v_i(t)}{2\sqrt{a_{max}b}} + \lambda(D_{i+1}(t) - D_i(t)), \tag{27}$$

$$\Delta v_i(t) = v_{i-1}(t) - v_i(t), \tag{28}$$

$$a_i(t + \Delta t) = a_{max}\left[1 - \left(\frac{v_i(t)}{v_f}\right)^4 - \left(\frac{S_i(t)}{D_i(t)}\right)^2\right]. \tag{29}$$

where $T$ represents the safe headway; $\Delta v_i(t)$ represents the relative speed of the vehicle $i$ to the preceding vehicle $i-1$ at time $t$; $a_{max}$ represents the maximum acceleration; $b$ represents the comfortable deceleration; $\lambda$ is the coefficient weight of balance spacing; $a_i(t + \Delta t)$ represents the acceleration of the vehicle at time $t + \Delta t$.

Table 5. The parameters of the BDBM model.

| Parameters | Value | Unit |
|---|---|---|
| $v_f$ | 33.3 | m/s |
| $T$ | 2.5 | s |
| $a_{max}$ | 1 | m/s² |
| $b$ | 2 | m/s² |
| $d_0$ | 2 | m |
| $\lambda$ | 0.5 | -- |
| $L$ | 5 | m |

By comparing the IDM and BDBM models, it can be seen that the BDBM model adds spacing difference and sets a weight coefficient $\lambda$. The BDBM model can be transformed into an IDM when the current spacing is balanced. According to Yi et al. (2022), the parameter values of the BDBM model were determined, as shown in Table 5.

### 4.5. Combination spacing strategy for mixed platoon

When each CAV adopts the same control strategy (i.e., CTG, VTG, BS strategy), it can better ensure vehicle stability because it maintains a relatively large following spacing from the preceding vehicle. This approach weakens the possibility of amplifying minor traffic disturbances during longitudinal propagation, thereby suppressing traffic oscillations. However, increasing the following spacing is not conducive to improving traffic capacity and may provide opportunities for vehicles on adjacent lanes to change lanes laterally, thereby exacerbating the instability of traffic flow. Based on this, this paper proposes multiple combination spacing strategies, where adjacent CAVs can form a flexible platoon through V2V communication. This method allows the following vehicles in the platoon to adopt a control strategy with a smaller expected following spacing, while the leading vehicle in the platoon does the opposite to achieve a buffering effect, as shown in Table 6. The CTG-CS strategy has been proposed by Zheng et al. (2023). Theoretically, this CAV platoon combination control strategy can balance stability and traffic efficiency compared to a spacing control strategy.

Table 6. Control strategies of CAV platoon.

| Index | LV | PV | Control strategy |
|---|---|---|---|
| 1 | CTG | CTG | |
| 2 | VTG1 | VTG1 | Single strategy |
| 3 | VTG2 | VTG2 | |
| 4 | BS | BS | |
| 5 | CTG | CS | |
| 6 | VTG1 | CTG | |
| 7 | VTG1 | CS | Combination strategy |
| 8 | VTG2 | CTG | |
| 9 | VTG2 | CS | |
| 10 | BS | CS | |

It is worth noting that the application mode of the BS-CS combination strategy is different from other combination strategies due to the need to use rear vehicle information. This strategy is based on the CAV platoon as a unit. As the following vehicles in the platoon adopt the CS strategy, the CAV platoon can be regarded as an "extended vehicle" using the BS strategy. Its adjacent rear vehicle may be another "extended vehicle" or an HV vehicle, as shown in Fig. 7. The advantage of this setting is that the essence of the BS-CS strategy is still the BS-BS strategy, which can ensure the stability of the vehicle when the penetration rate of CAVs is 100%. The disadvantage is that it requires high communication requirements.

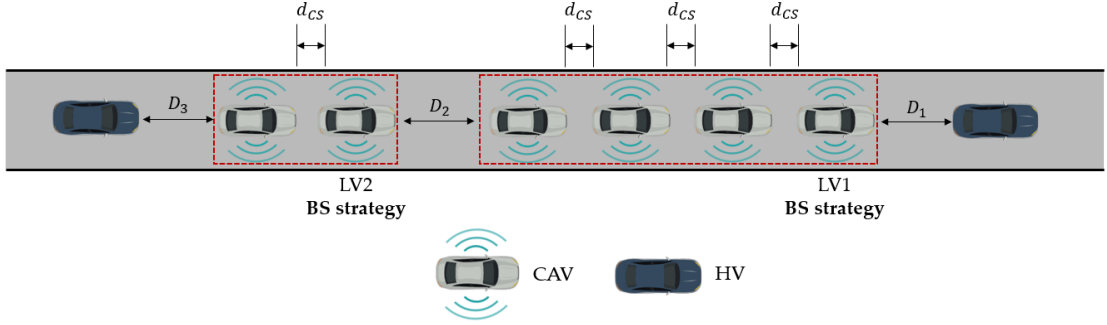

Fig. 7. The BS-CS Strategy.

**5. Fuel consumption and emissions**

5.1. Fuel consumption model

A vehicle's fuel consumption is directly related to the engine output power. In existing research, vehicle-specific power (VSP, unit: kW/ton) is widely used in fuel consumption modeling. For typical light vehicles, assuming a road slope of 0, the expression for VSP is

$$VSP = v \cdot [1.1a + 0.132] + 0.000302v^3, \qquad (30)$$

where $v$ is the vehicle speed (m/s); $a$ is the vehicle acceleration (m/s$^2$).

The absolute fuel consumption rate calculated through VSP is influenced mainly by engine size, fuel type, and vehicle mass. The normalized fuel consumption rate (NFR, unit: g/s) proposed by Song and Yu (2009) can avoid this impact. The regression relationship between NFR and VSP is represented by Eq. (31), where VSP values are positive. When VSP is negative, NFR takes 1.

$$NFR = 1.71 \cdot VSP^{0.42}. \qquad (31)$$

By substituting the speed and acceleration into Eq. (31), the heatmap related to the normalized fuel consumption rate can be obtained in Fig. 8.

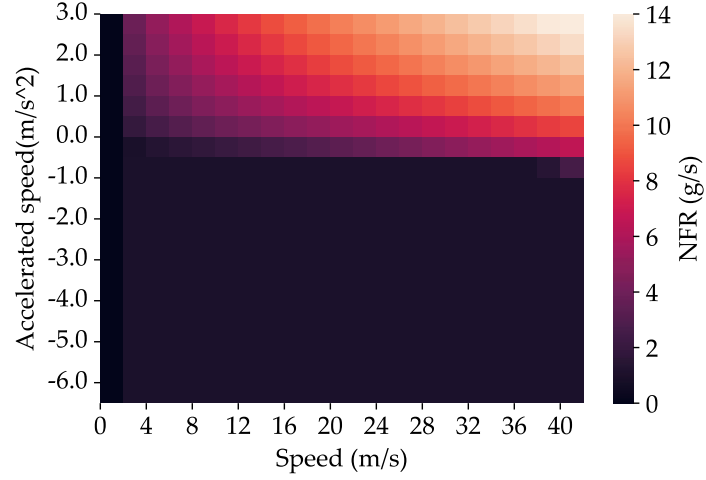

Fig. 8. NFR v.s. speed and acceleration.

Fig. 8 shows that the fuel consumption model emphasizes the vehicle does not consume fuel when decelerating but only consumes fuel when accelerating and moving at a constant speed. Moreover, the fuel consumed per second is proportional to acceleration and speed.

The calculation formula for the average NFR of a specific transportation network is shown in Eq. (32), which is the sum of the products of the NFR corresponding to each VSP and its time proportion fraction.

$$\overline{NFR} = \sum_j time\ fraction_j \cdot NFR_j, \tag{32}$$

where $j$ is the VSP number.

The average NFR is the normalized fuel consumption per unit time, which can be divided by the network's average driving speed (km/h) to obtain the normalized fuel consumption per unit distance. It is represented by the normalized fuel factor (NFF, unit: g/km), as shown in Eq. (33).

$$\overline{NFF} = 3600 \times \frac{\overline{NFR}}{\bar{v}}. \tag{33}$$

where $\bar{v}$ is the average speed of the vehicle (m/s).

If we assume the traffic flow is in equilibrium, the following line graph about the normalized fuel coefficient is calculated by substituting the equilibrium velocity values.

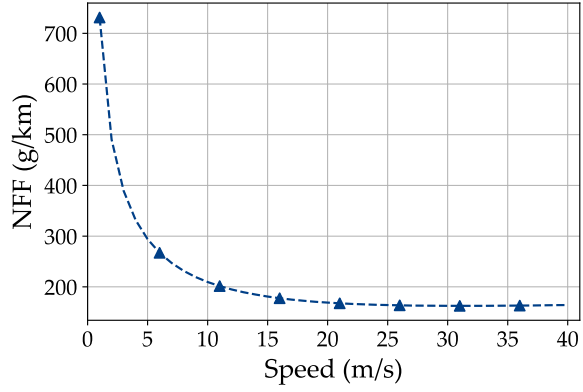

Fig. 9. The equilibrium state of the fuel consumption model.

Fig. 9 presents that NFF shows a sharply decreasing trend and then gradually stabilizes with the increase of equilibrium speed. At low speeds, the fuel consumption is higher than at high speeds. This is because although higher speeds consume more fuel per unit of time, it takes less time to travel the same distance. Therefore, using the NFF index to evaluate the fuel consumption level of traffic flow should be more realistic. Meanwhile, according to the fuel consumption model, to maintain the fuel consumption at a low level in equilibrium and save energy, the overall speed of traffic flow should be increased.

5.2. Emissions model

Int Panis et al. (2006) established the following vehicle emission model based on nonlinear multiple regression techniques,

$$E_i(t) = \max[0, f_1 + f_2 v_i(t) + f_3 v_i(t)^2 + f_4 a_i(t) + f_5 a_i(t)^2 + f_6 v_i(t) a_i(t)]. \tag{34}$$

where $E_i(t)$ is the emission amount (g/s) of vehicle $i$ at time $t$. $f_1$ to $f_6$ are the specific emission coefficients for each pollutant, with values shown in Table 7.

Table 7. The coefficient of pollutant emission.

| Pollutant emission | | $f_1$ | $f_2$ | $f_3$ | $f_4$ | $f_5$ | $f_6$ |
|---|---|---|---|---|---|---|---|
| $CO_2$ | — | | 5.53e-01 | 1.61e-01 | -2.89e-03 | 2.66e-01 | 5.11e-01 | 1.83e-01 |
| $NO_x$ | $a \geq -0.5 m/s^2$ | 6.19e-04 | 8.00e-05 | -4.03e-06 | -4.13e-04 | 3.80e-04 | 1.77e-04 |
|  | $a < -0.5 m/s^2$ | 2.17e-04 | 0 | 0 | 0 | 0 | 0 |
| VOC | $a \geq -0.5 m/s^2$ | 4.47e-03 | 7.32e-07 | -2.87e-08 | -3.41e-06 | 4.94e-06 | 1.66e-06 |
|  | $a < -0.5 m/s^2$ | 2.63e-03 | 0 | 0 | 0 | 0 | 0 |
| PM | — | 0 | 1.57e-05 | -9.21e-07 | 0 | 3.75e-05 | 1.89e-05 |

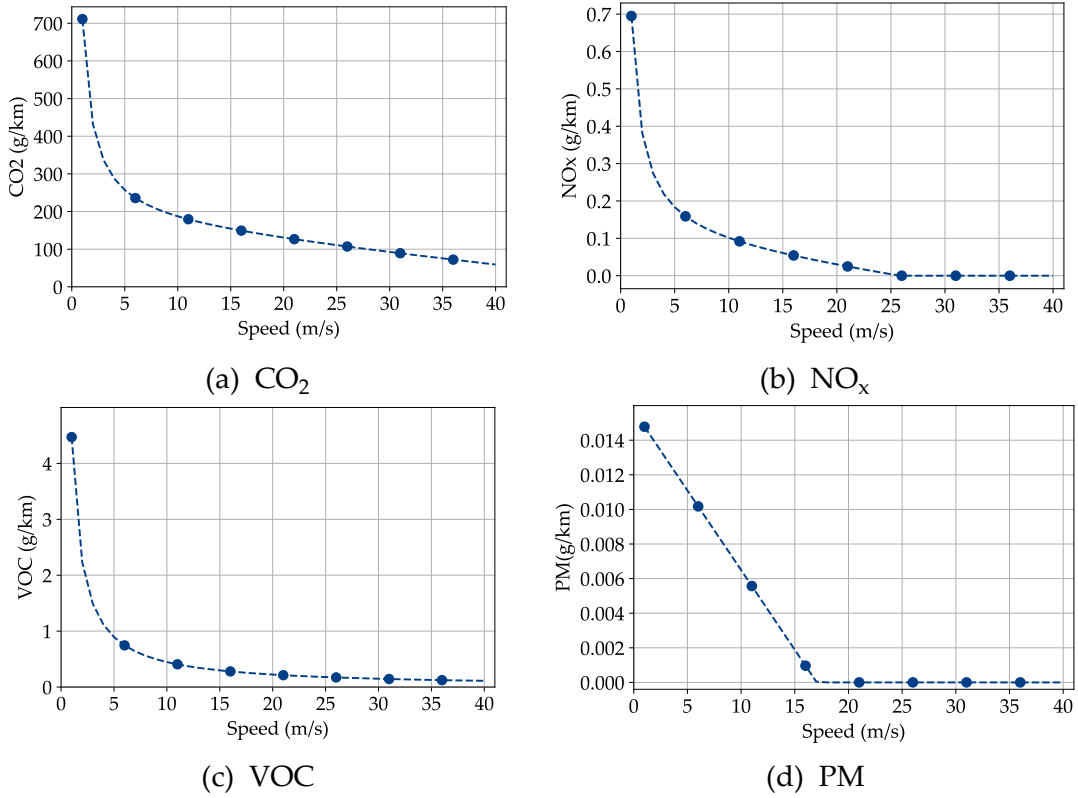

(a) $CO_2$

(b) $NO_x$

(c) VOC

(d) PM

Fig. 10. The emission in equilibrium state.

Similarly, pollutant emission evaluation indicators based on unit distance are used. The traffic flow is in a state of equilibrium, and the equilibrium velocity value is substituted and calculated, as shown in Fig. 10. From Fig. 10, it can be seen that the emissions of the four pollutants decrease with increasing equilibrium velocity, among which the emissions of NOx, VOC, and PM per kilometre gradually approach or equal to zero with increasing velocity. When the speed exceeds 20 m/s, the PM emissions remain constant at 0. When the vehicle speed exceeds 25 m/s, the NOx emissions remain constant at 0. Under low-speed conditions, the decrease in PM emissions is linear, while the other three are curvilinear decreases. When a vehicle travels at low and constant speed, it emits many pollutants, causing air pollution. To reduce the pressure of vehicle exhaust emissions on the environment, it is necessary to ensure that the traffic flow can run at a higher speed when in equilibrium.

5.3. Numerical simulation

5.3.1 Simulation settings

This paper conducts numerical simulation experiments on a single lane based on Python 3.8. The simulation experiment aims to obtain the trajectory of traffic flow under different density conditions. According to Zhou et al. (2021), the ring road simulation experiment is conducted in this paper.

The length in the ring road simulation experiment is set to 1000 m. Vehicles are evenly distributed on the ring road in the initial state, and their initial velocity and acceleration values are 0. Based on platoon intensity $O = 1$, gather all CAVs to form a long platoon. To be more realistic, the maximum speed of the vehicle is set to 33.3 m/s, the maximum acceleration is 1 m/s$^2$ (Wang et al., 2019), the minimum acceleration is

-5 m/s$^2$ (Jiang et al., 2021), the penetration rates of CAVs are set to 0, 0.2, 0.4, 0.6, 0.8, and 1. The traffic density range is set from 5 veh/km to 100 veh/km with a step of 5 veh/km. The setting aims to evaluate the performance of the proposed strategy in terms of fuel consumption and pollutant emissions under different penetration rates of CAVs. The total simulation duration is 3600 s, and the simulation step is 0.1 s.

5.3.2 The impact of different spacing combination strategies on fuel consumption

The traffic flow speed and acceleration data of the simulation experiment for half an hour are obtained to calculate the average fuel consumption of the mixed traffic flow. Then, the average fuel consumption corresponding to each control strategy under different penetration rates of CAVs is drawn, as shown in Fig. 11.

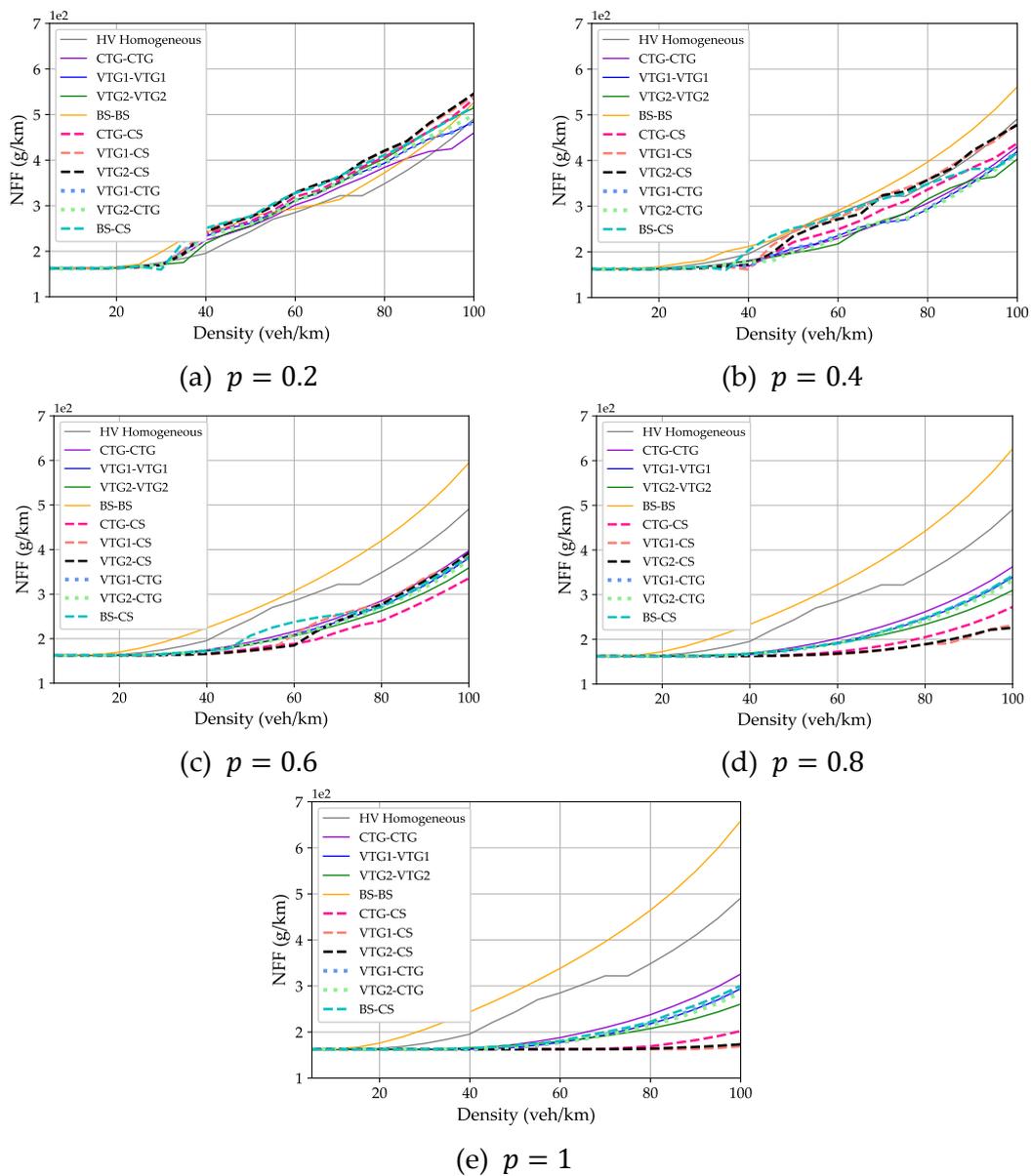

Fig. 11. The average fuel consumption with different control strategies and penetration rates of CAVs.

To facilitate comparison with HVs homogeneous traffic flow, separate curves with $p = 0$ were drawn using gray solid lines in each subplot. As traffic density increases, the average fuel consumption corresponding to all control strategies gradually increases. As the penetration rate of CAVs increases, the fuel consumption corresponding to the BS-BS strategy gradually increases, while the other 9 strategies are the opposite. When the penetration rate of CAVs is 100%, the VTG1-CS and VTG2-CS strategies correspond to the slightest increase, and the curve is almost flat.

It was found that the impact of the penetration rate of CAVs on the average fuel consumption corresponding to each control strategy varies under different traffic densities. For the convenience of analysis, this section takes three traffic density values of low, medium, and high (corresponding to 15 veh/km, 55 veh/km, and 95 veh/km, respectively) and plots NFT v.s. the penetration rate of CAVs for each control strategy, as shown in Fig. 12.

As shown in Fig. 12, when the traffic density is 15 veh/km, the difference in average fuel consumption values corresponding to the 10 control strategies is insignificant and almost does not change with the change of penetration rate of CAVs. This is because vehicles travel at extremely high speeds under low-density conditions. Moreover, according to Fig. 9, when the equilibrium speed of the vehicle is higher than 20 m/s, the fuel consumption remains almost unchanged and reaches the lowest level.

When the traffic density is 55 veh/km, the average fuel consumption corresponding to the BS-BS strategy shows a trend of "increasing first, then decreasing, and then increasing again". The first increase in fuel consumption was due to the instability of the mixed traffic flow and significant fluctuations in vehicle speed when the penetration rate was 20%. The subsequent reduction in fuel consumption is because the traffic flow is already stable when the penetration rate reaches 40%. The curve shows a linear upward trend when the penetration rate is greater than or equal to 60%. This means that the BS-BS strategy will cause the vehicle to consume more fuel. For the other 9 control strategies, there is a trend of "increasing first and then decreasing", and the rate of decline slows down in the later stage. The first increase was also affected by the instability of mixed traffic flow, during which the BS-CS strategy had the highest fuel consumption of approximately 300 g/km. Subsequently, with the addition of more CAVs, the traffic flow gradually stabilized, and the speed increased further, resulting in a decrease in overall fuel consumption. When the penetration rate is greater than 60%, the fuel consumption corresponding to the VTG1-CS, VTG2-CS, and CTG-CS strategies is relatively the lowest compared to other strategies. When the penetration rate is greater than or equal to 80%, the CTG-CTG strategy is the highest. From Fig. 12 (b), it can also be observed that the VTG2-VTG2 strategy is more fuel-efficient than other strategies within the range of penetration rate less than or equal to 40%, while the BS-CS, VTG1-CS, and VTG2-CS strategies, constrained by unstable traffic flow factors, are more fuel-efficient within the range above.

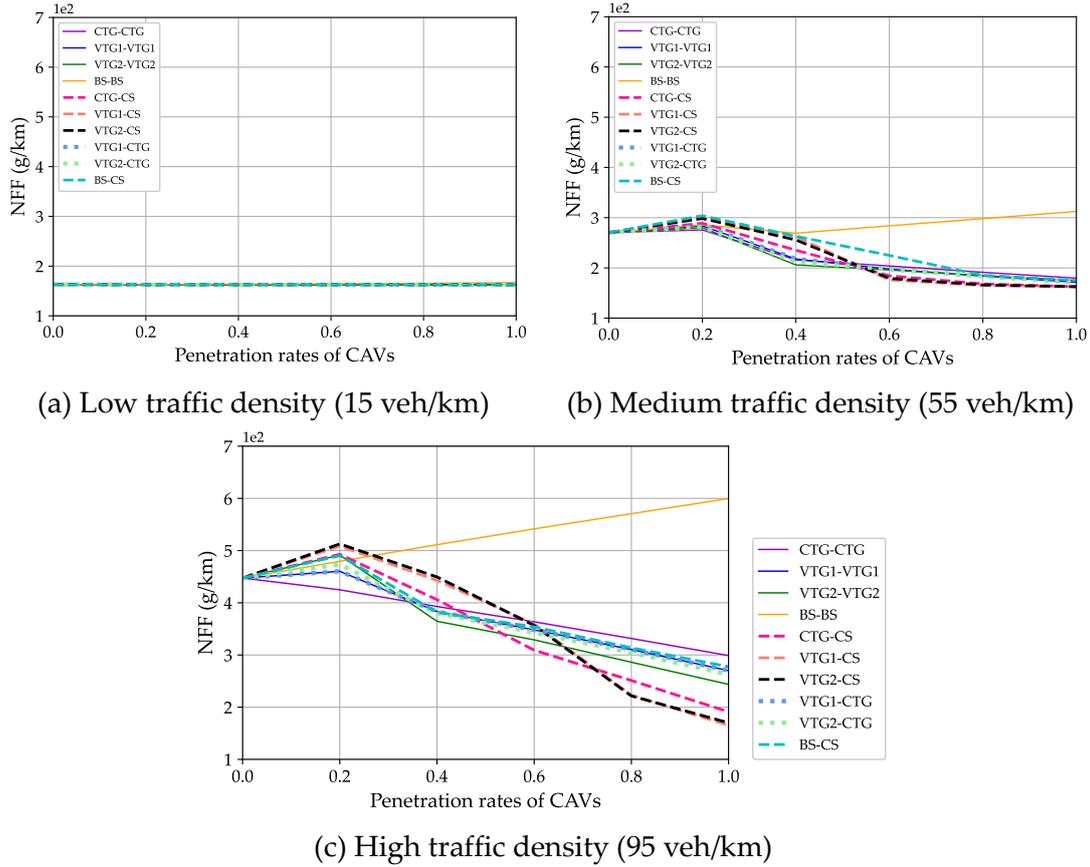

(a) Low traffic density (15 veh/km)

(b) Medium traffic density (55 veh/km)

(c) High traffic density (95 veh/km)

Fig. 12. Average fuel consumption corresponding to each control strategy with different traffic density.

When the traffic density is 95 veh/km, there are significant differences among the control strategies. For the BS-BS strategy, the overall speed of traffic flow is relatively stable under different CAV penetration rates. As the penetration rate of CAVs increases, the average speed corresponding to the BS-BS strategy gradually decreases, and the average fuel consumption value shows a linear upward trend. On the contrary, the CTG-CTG strategy shows a linear downward trend in average fuel consumption, and is most energy-efficient at low penetration rates of CAVs. This indicates that this strategy can effectively stabilize mixed traffic flow under high-density and low penetration rate conditions. As for the other 8 strategies still show a trend of "increasing first and then decreasing", and the rate of decline slows down in the later stage.

Similarly, constrained by unstable traffic flow factors, fuel consumption peaks at a CAV penetration rate of 20%, with VTG2-CS and VTG1-CS strategies being the most fuel-efficient, with an average fuel consumption value exceeding 500 g/km. As the penetration rate of CAV increases, traffic flow gradually stabilizes, and speed increases, decreasing overall fuel consumption. When the penetration rate is greater than or equal to 80%, all eight strategies are lower than the CTG-CTG strategy, with the VTG1-CS and VTG2-CS strategies corresponding to the lowest fuel consumption, followed by the CTG-CS strategy. From Fig. 12 (c), it can also be observed that when the penetration rate is 40%, the fuel consumption of the VTG2-VTG2 strategy is the lowest, while the CTG-CS strategy is slightly higher than the CTG-CTG strategy. When the penetration rate is 60%, the CTG-CS strategy is the lowest, and the CTG-CTG strategy is the highest, except for the BS-BS strategy.

5.3.3 The impact of different combination strategies on pollutant emissions

The traffic flow velocity and acceleration data from the simulation experiment are obtained to calculate the average emissions of four pollutants in the mixed traffic flow. The results and related analysis are as follows.

(1) $CO_2$

A line graph of traffic flow CO2 emissions for each control strategy under different penetration rates of CAVs is drawn, as shown in Fig. 13.

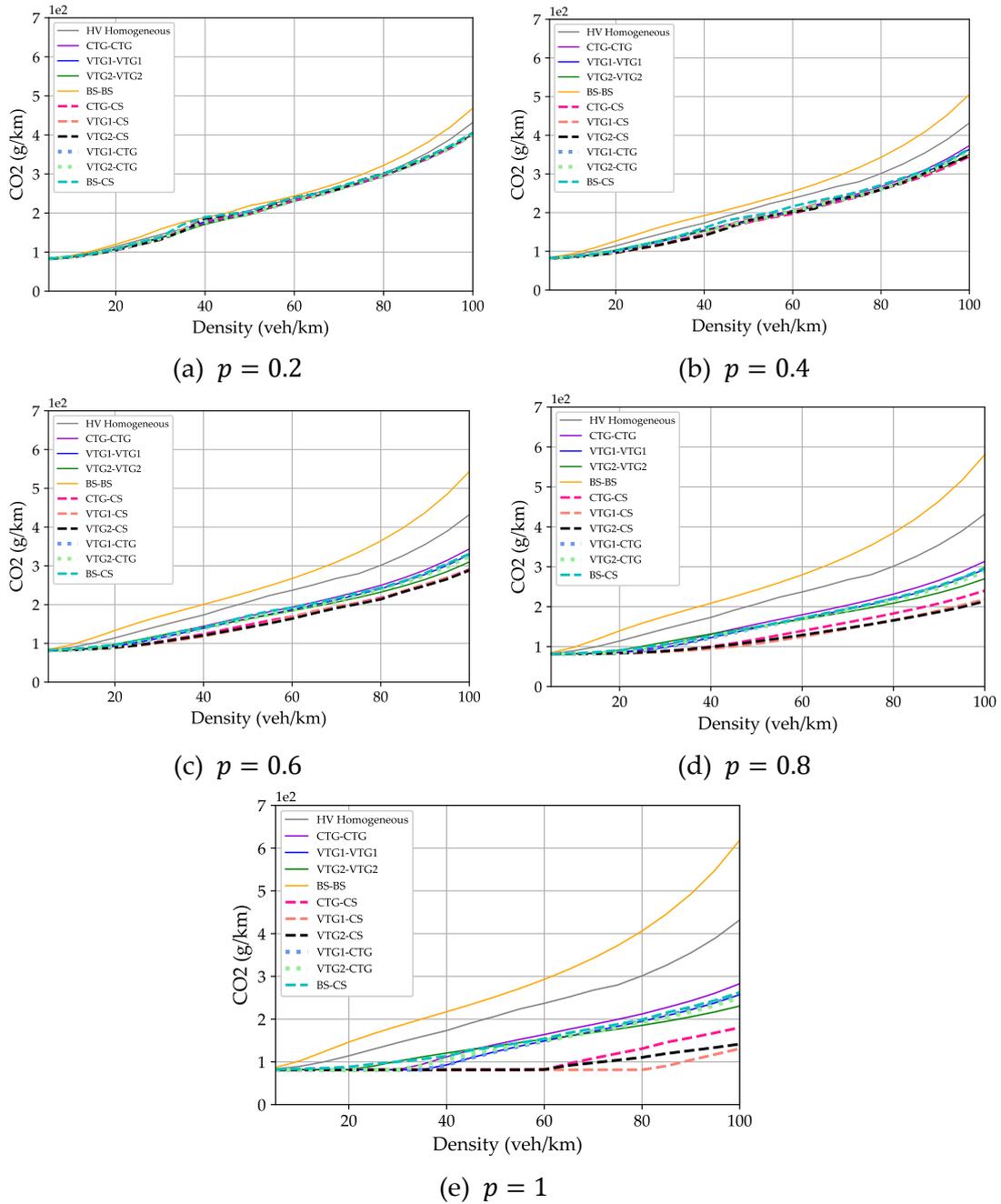

Fig. 13. The average CO2 emissions corresponding to each control strategy under different penetration rates of CAVs.

For the convenience of comparing with HV homogeneous traffic flow, separate curves with $p = 0$ were drawn using gray solid lines in each subfigure. The same applies to the remaining three pollutant emission-related line graphs, which will not be repeated in the following text. As traffic density increases, the average $CO_2$ emissions of all control strategies gradually increase. As the penetration rate of CAVs increases, the $CO_2$ emissions corresponding to the BS-BS strategy also gradually increase, while the other 9 strategies are the opposite, and the corresponding curve differences become more evident. When the penetration rate is 100%, the increase corresponding to the VTG1-CS strategy is the least. Within the traffic density range of 0 to 60 veh/km, the increased values corresponding to the VTG1-CS, VTG2-CS, and CTG-CS strategies are about 0 g/km.

It can also be found that the impact of the penetration rate of CAVs on the average $CO_2$ emissions corresponding to each control strategy varies under different traffic densities. For the convenience of analysis, this section takes three traffic density values (corresponding to 15 veh/km, 55 veh/km, and 95 veh/km, respectively) for low, medium, and high, and draws the CAVs penetration rate v.s. CO2 emission line charts for each control strategy, as shown in Fig. 14.

As shown in Fig. 14, when the traffic density is 15 veh/km, the average $CO_2$ emissions corresponding to the BS-BS strategy increase with the penetration rates of CAVs, while the remaining 9 strategies are the opposite and have almost the same emissions. This is because, under low-density conditions, the average speed of traffic flow corresponding to the other 9 strategies is not significantly different and is greater than the average speed of HV homogeneous traffic flow, while the BS-BS strategy is lower than HV homogeneous traffic flow. Combining Fig. 10 (a), the $CO_2$ emissions will decrease with the increase of equilibrium velocity so that the above conclusion can be drawn.

When the traffic density is 55 veh/km, the average $CO_2$ emissions corresponding to the BS-BS strategy still show a linear upward trend. For the remaining 9 control strategies, the average $CO_2$ emissions will decrease with the increase of penetration rates of CAVs, and the rate of decrease will gradually increase. When the penetration rates of CAVs is greater than 40%, the $CO_2$ emissions corresponding to the VTG1-CS, VTG2-CS, and CTG-CS strategies are the lowest, and even at a penetration rate of 100%, the amount of $CO_2$ produced is equivalent to that of low-density. The remaining 6 strategies have relatively minor differences, but the CTG-CTG strategy is the highest. When the penetration rate of CAVs is less than or equal to 40%, the $CO_2$ emissions of these 9 strategies are not significantly different.

When the traffic density is 95 veh/km, there are significant differences in various control strategies, but the primary trend is the same as under medium density. Unlike the medium density, when the penetration rates of CAVs are more excellent than 60%, the $CO_2$ emissions under the CTG-CS strategy are higher than those under the VTG1-CS and VTG2-CS strategies. Moreover, when the penetration rates of CAVs are more excellent than 80%, the difference between the VTG1-CS strategy and the VTG2-CS strategy becomes significant, with the VTG1-CS strategy corresponding to slightly lower $CO_2$ emissions. For the six control strategies at the intermediate level, it can be found that when the penetration rates of CAVs are greater than 40%, the amount of $CO_2$ emitted from traffic flow under the VTG2-VTG2 strategy is the lowest.

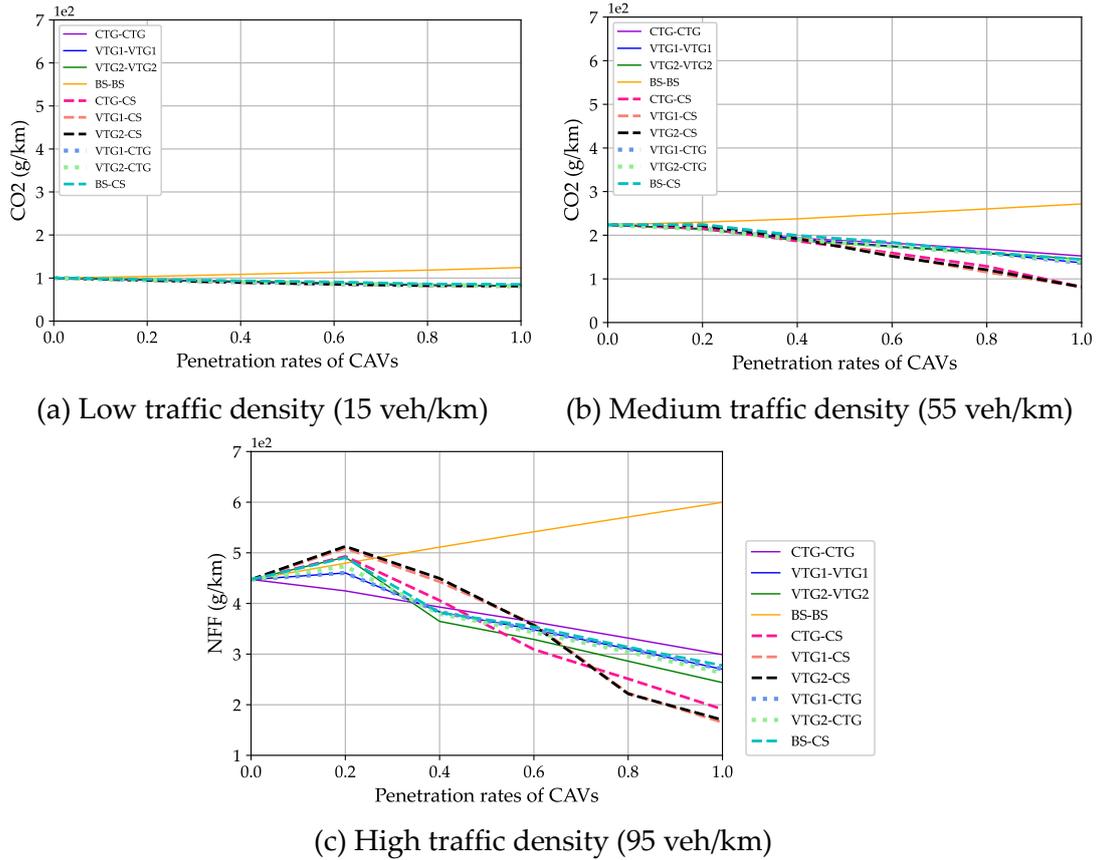

(a) Low traffic density (15 veh/km)
(b) Medium traffic density (55 veh/km)
(c) High traffic density (95 veh/km)

Fig. 14. The average $CO_2$ emissions corresponding to each control strategy under different traffic densities.

(2) $NO_x$

A line graph of traffic flow $NO_x$ emissions for each control strategy under different penetration rates of CAVs is drawn, as shown in Fig. 15. As traffic density increases, the average $NO_x$ emissions corresponding to all control strategies gradually increase. As the penetration rate of CAVs increases, the $NO_x$ emissions corresponding to the BS-BS strategy also gradually increase, while the other 9 strategies are the opposite, and the corresponding curve differences become more significant. When the penetration rate of CAVs is 100%, the VTG1-CS strategy corresponds to the slightest increase, followed by the VTG2-CS strategy. Within the traffic density range of 0 to 70 veh/km, the increase values corresponding to the VTG1-CS, VTG2-CS, and CTG-CS strategies are approximately 0 g/km.

It can also be found that under different traffic densities, the sensitivity of average $NO_x$ emissions v.s. CAV penetration rate with varying strategies of control. For the convenience of analysis, the three traffic densities, low, medium, and high (corresponding to 15 veh/km, 55 veh/km, and 95 veh/km, respectively) were also taken. Then, the CAV penetration rate v.s. $NO_x$ emission line charts corresponding to each control strategy were plotted, as shown in Fig. 16.

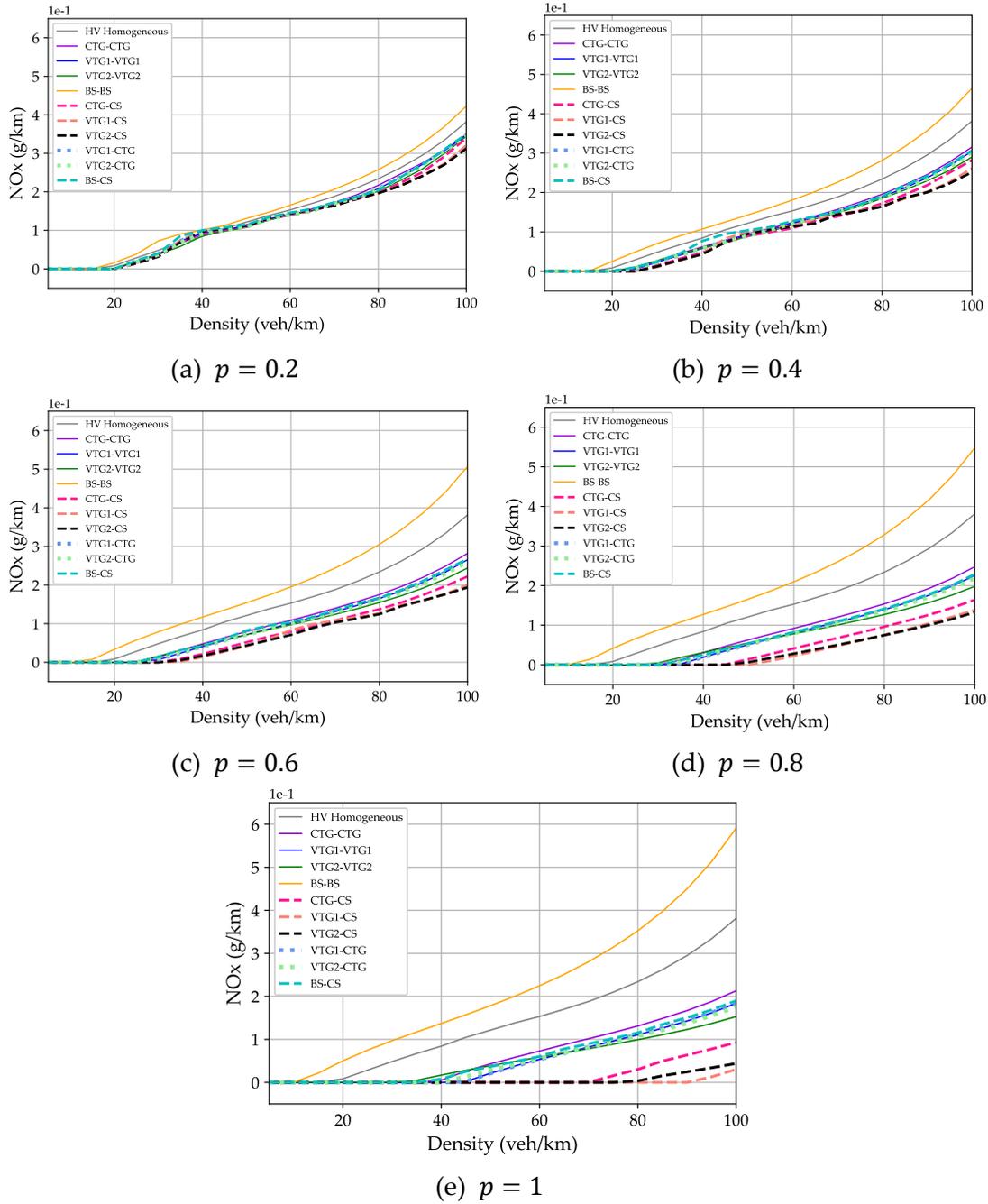

(a) $p = 0.2$
(b) $p = 0.4$
(c) $p = 0.6$
(d) $p = 0.8$
(e) $p = 1$

Fig. 15. The average NOx emissions corresponding to each control strategy under different penetration rates of CAVs.

As shown in Fig. 16, when the traffic density is 15 veh/km, the average NOx emissions corresponding to the BS-BS strategy show a significant upward trend under a high penetration rate, while the curves corresponding to the remaining 9 strategies are almost flat. This is because, in low-density traffic scenarios, the average speed of traffic flow corresponding to the other 9 strategies is similar and maintains a relatively high level, surpassing the average speed of HV homogeneous traffic flow. However, the BS-BS strategy performed poorly, as its average traffic flow speed did not reach the HV homogeneous traffic flow level. Combining Fig. 10 (b), the NOx emissions will decrease with the increase of equilibrium velocity, and when the velocity is greater than 25 m/s, the emissions are about 0 g/km. When the traffic density is 55 veh/km, the average NOx emissions corresponding to the BS-BS strategy still show a linear upward trend, with a maximum value of about 0.2 g/km. The average NOx emissions of the remaining 9 control strategies will decrease as the penetration rate increases. When the penetration rate of CAVs reaches 100%, the NOx emissions corresponding to the VTG1-CS, VTG2-CS, and CTG-CS strategies are approximately 0 g/km. The remaining 6 strategies have relatively minor differences, but it can be seen that the CTG-CTG strategy is the highest. When the traffic density is 95 veh/km, there are significant differences among the control strategies. The trend of the BS-BS strategy still shows a linear increase, while other strategies decrease with the penetration rate increase of CAVs. The decline rate of VTG1-CS, VTG2-CS, and CTG-CS strategies shows a significant increase under high penetration rate, especially the VTG1-CS strategy. As for the remaining six strategies regarding reducing NOx emissions, the VTG2-VTG2 strategy is the best, and the CTG-CTG strategy is the worst, but overall, it still falls short of the three strategies above.

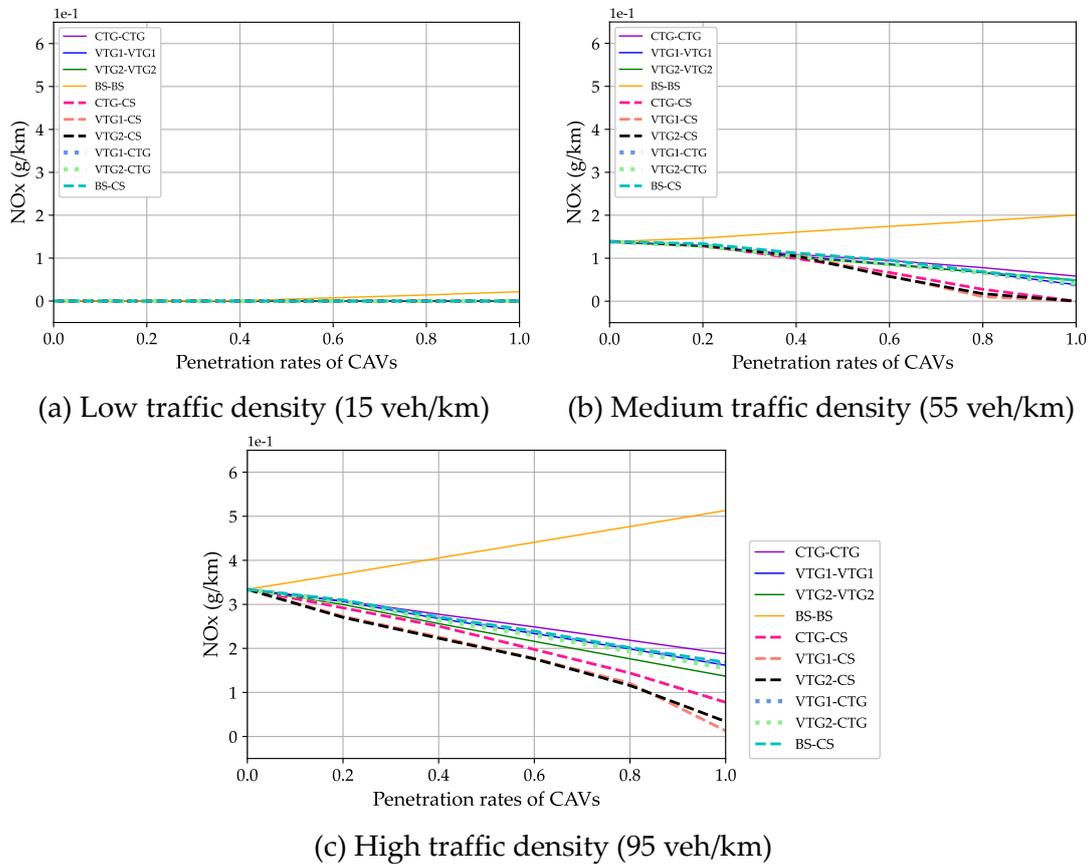

(a) Low traffic density (15 veh/km)  (b) Medium traffic density (55 veh/km)

(c) High traffic density (95 veh/km)

Fig. 16. The average NOx emissions corresponding to each control strategy under different traffic density.

(3) VOC

A line graph of the average VOC emissions corresponding to the control strategies under different penetration rates is drawn, as shown in Fig. 17. As traffic density increases, the average VOC emissions of traffic flow corresponding to all control strategies gradually increase. As the penetration rate of CAVs increases, the VOC emissions corresponding to the BS-BS strategy also gradually increase, while the other 9 strategies are the opposite, and the corresponding curve differences become more significant. When the penetration rate of CAVs is 100%, the VTG1-CS and VTG2-CS strategies correspond to the slightest increase, about 0 g/km, followed by the CTG-CS strategy, but the increase is still less than 0.5 g/km.

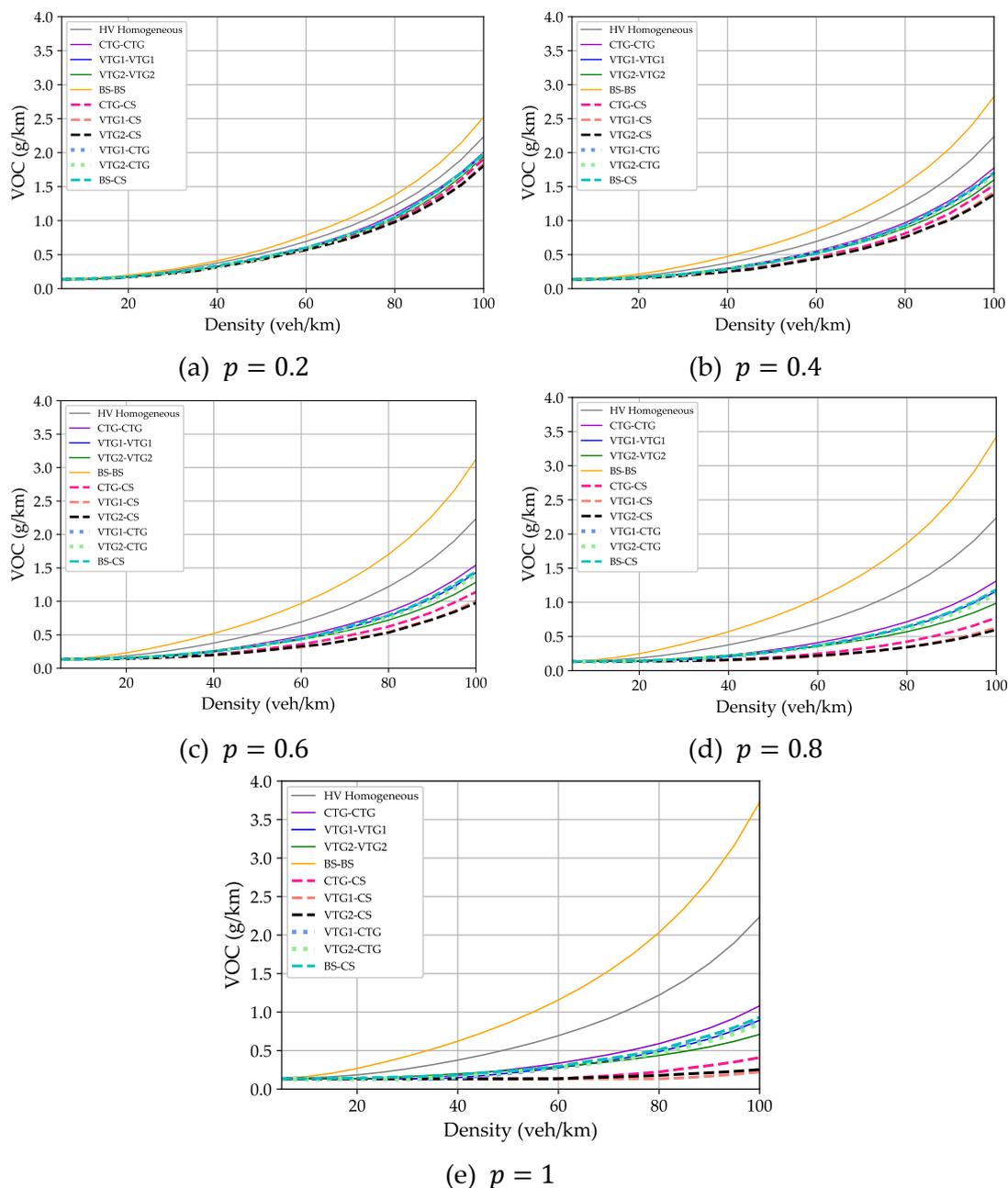

Fig. 17. The average VOC emissions corresponding to each control strategy under different penetration rates of CAVs.

It can also be found that implementing different control strategies under different traffic density conditions will lead to significant differences in the sensitivity of average VOC emissions to CAV penetration rate. Similarly, select density values for low, medium, and high levels, namely 15 veh/km, 55 veh/km, and 95 veh/km, respectively. Based on these three traffic density values, a line graph of CAV penetration rate v.s. VOC emissions corresponding to each control strategy are drawn, as shown in Fig. 18.

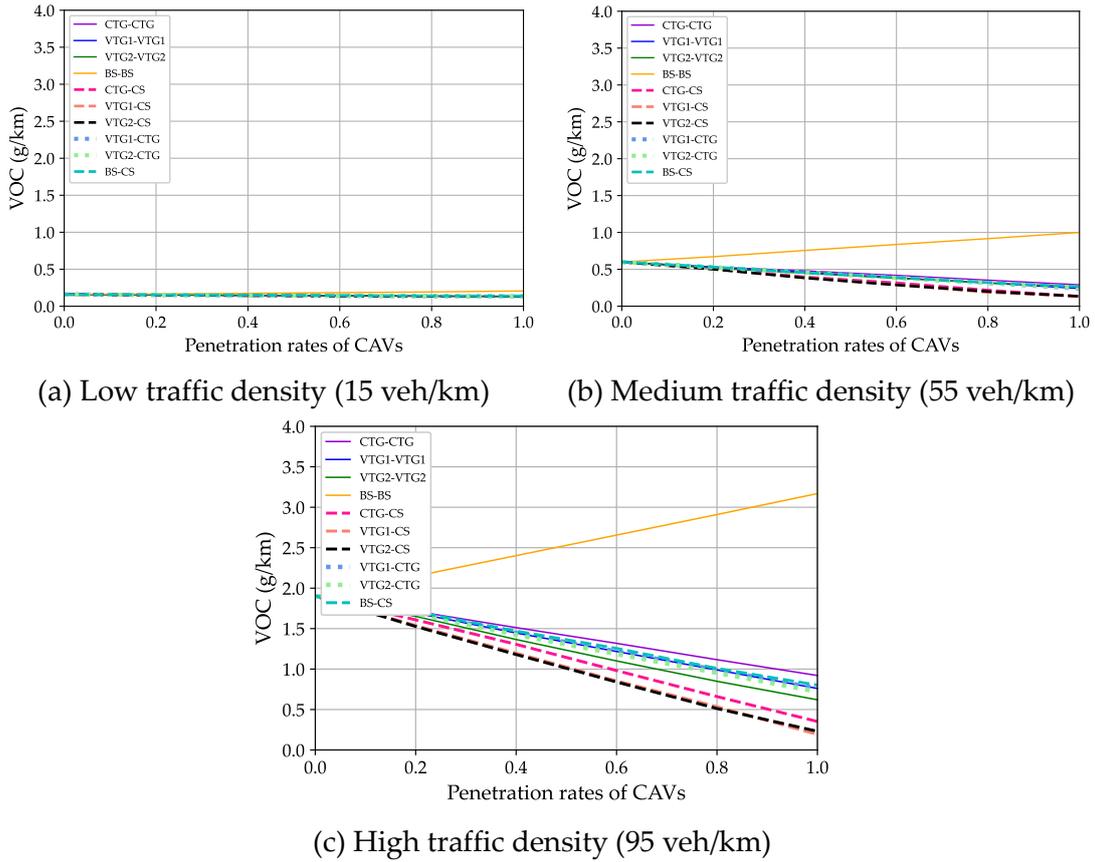

(a) Low traffic density (15 veh/km)  (b) Medium traffic density (55 veh/km)

(c) High traffic density (95 veh/km)

Fig. 18. The average VOC emissions corresponding to each control strategy under different traffic densities.

Fig. 18 shows that when the traffic density is 15 veh/km, there is little difference in the average VOC emissions corresponding to the 10 control strategies. This is also because, under low-density conditions, vehicles can travel at extremely high speeds. Fig. 10 displays that when the equilibrium speed of the vehicle is higher than 20 m/s, the change in VOC emissions is relatively small, almost reaching the lowest level.

When the traffic density is 55 veh/km, the average VOC emissions corresponding to the BS-BS strategy show a linear upward trend, with a maximum value of about 1 g/km. The remaining 9 control strategies' average VOC emissions are inversely proportional to the penetration rate CAVs, among which VTG1-CS, VTG2-CS, and CTG-CS strategies perform better in reducing vehicle VOC emissions. When the traffic density is 95 veh/km, there are significant differences in the control strategies. The BS-BS strategy shows a linear upward trend, and its VOC emissions remain stable. However, all other strategies decrease with the increase of penetration rate of CAVs, among which VTG1-CS and VTG2-CS strategies have the highest decline rate and correspondingly the lowest VOC emissions, followed by CTG-CS strategy, VTG2-VTG2 strategy, and CTG-CTG strategy.

(4) PM

A line graph of the average PM emissions corresponding to the control strategies under different penetration rates of CAVs is drawn, as shown in Fig. 19. Generally speaking, the increase in traffic density has led to an upward trend in the average VOC emissions under all control strategies. Meanwhile, with the rise in the penetration rate of CAVs, the VOC emissions under the BS-BS strategy are also increasing, while the other 9 strategies show the opposite trend, and the differences between their curves become increasingly significant. The changes in VTG1-CS, VTG2-CS, and CTG-CS strategies are more drastic. When the penetration rate of CAVs is 100%, and the density is less than or equal to 80 veh/km, the corresponding PM emissions are all 0 g/km. This fully demonstrates that these three strategies perform outstandingly in reducing PM pollutant emissions, especially the combination of the two VTG strategies, and the CS strategy has the best effect.

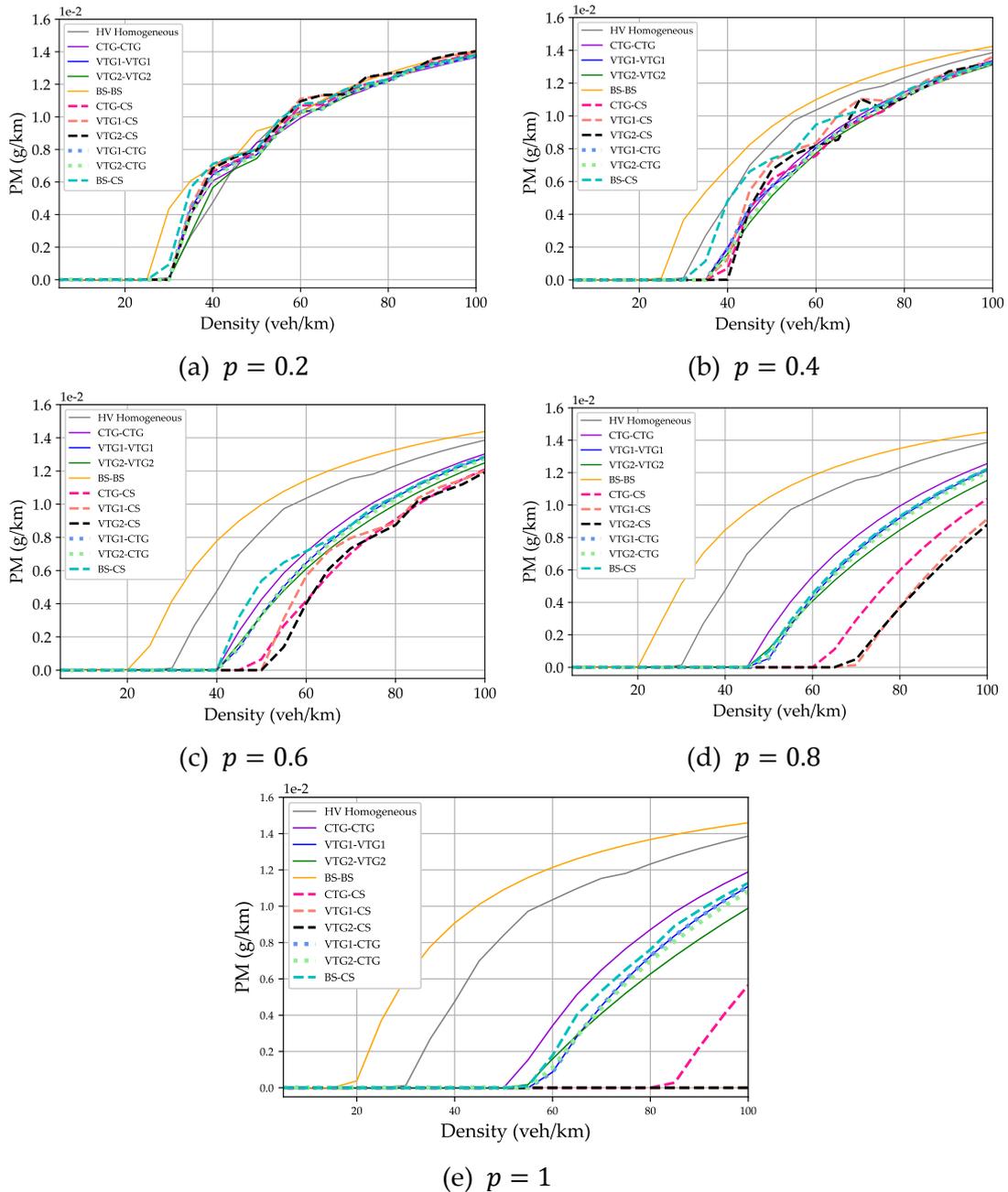

Fig. 19. The average PM emissions corresponding to each control strategy under different penetration rates of CAVs.

Similarly, under different traffic densities, the increase in penetration rates has varying degrees of impact on the average PM emissions under different control strategies. To simplify the analysis process, this paper selected three density values:

low (15 veh/km), medium (55 veh/km), and high (95 veh/km), and plotted line graphs of CAV penetration rate and PM emissions under different control strategies for these density values, as shown in Fig. 20.

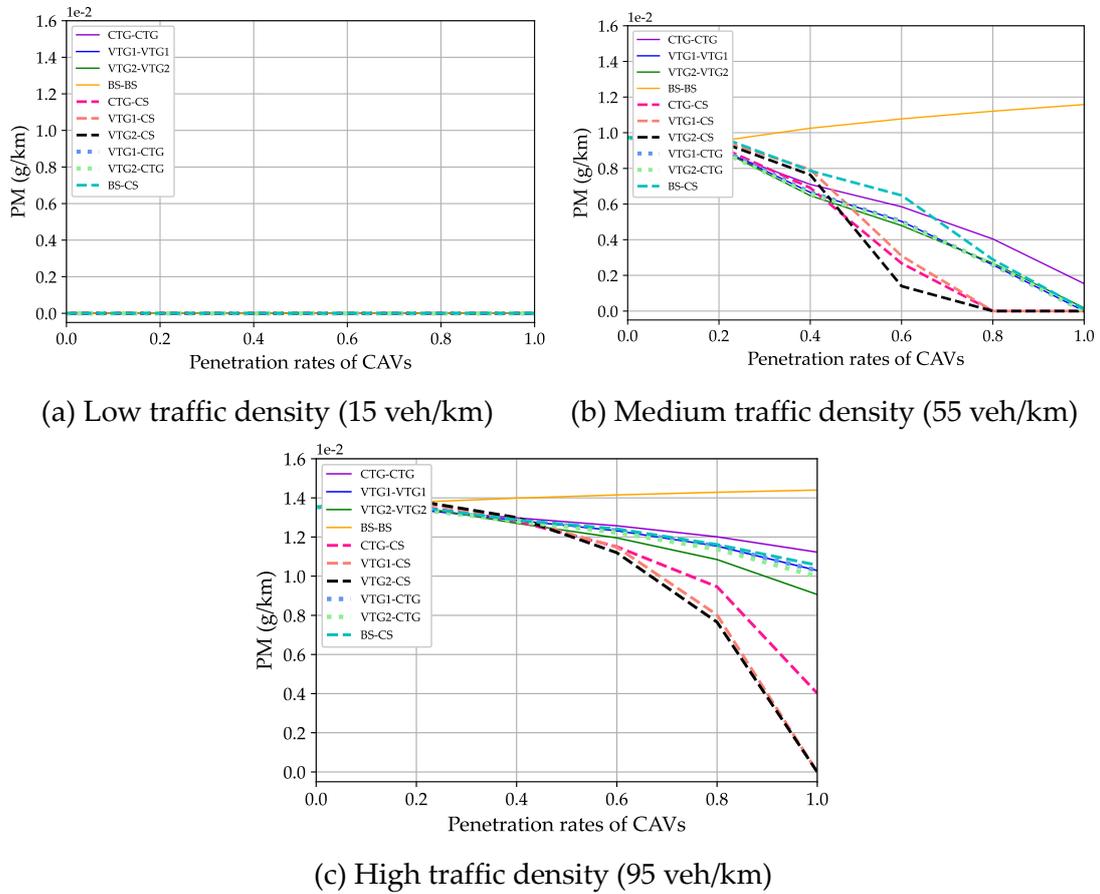

(a) Low traffic density (15 veh/km)　　(b) Medium traffic density (55 veh/km)

(c) High traffic density (95 veh/km)

Fig. 20. The average PM emissions corresponding to each control strategy under different traffic densities.

As can be seen from Fig. 20, when the traffic density is 15 veh/km, the difference in PM emissions corresponding to all control strategies is very insignificant and almost does not change with the change of penetration rates of CAVs, which is about 0 g/km. This is mainly due to the ability of vehicles to travel smoothly at extremely high speeds in low-density traffic environments, effectively suppressing PM emissions. When the traffic density is 55 veh/km, the overall trend of the average PM emissions corresponding to the BS-BS strategy increases with the penetration rates of CAVs, while the other strategies show the opposite trend. Under low penetration rates of CAVs, mixed traffic flows implementing BS-BS, BS-CS, VTG1-CS, and VTG2-CS strategies have relatively high PM emissions. However, when the penetration rates of CAVs increase from 40% to 60%, the line corresponding to the VTG2-CS strategy decreases the most, followed by the CTG-CS and VTG1-CS strategies. When the penetration rate is 60%, the PM emissions under the VTG2-CS strategy are less than 0.002 g/km, and the CTG-CS and VTG1-CS strategies are less than 0.004 g/km. When the penetration rates of CAVs are greater than or equal to 80%, the PM emissions corresponding to the above three strategies are approximately 0 g/km. Under the penetration rates of CAVs of 100%, except for the BS-BS and CTG-CTG strategies, all other strategies are about 0 g/km. When the traffic density is 95 veh/km, although there is a slight upward trend in some strategies at low penetration rates, overall, the primary trend of each control strategy is consistent with that at medium traffic density. Similarly, under high penetration rates of CAVs, there are significant differences in various control strategies, among which VTG2-CS and VTG1-CS strategies have the best effect in reducing PM emissions, followed by CTG-CS strategy, and then VTG2-VTG2 strategy. The BS-BS strategy is the worst, followed by the CTG-CTG strategy. The remaining strategy differences are minimal. The other four strategies are not significantly different from the VTG1-VTG1 strategy.

## 6. Conclusions and Future Work

This paper proposes multiple spacing combination strategies for the CAV platoon. Through simulation experiments, the traffic flow under different traffic conditions is simulated. The results are used to analyze the impact of different spacing combination strategies on traffic flow fuel consumption and pollutant emissions. The following conclusions are drawn.

(1) With the increase in traffic density, the average speed of traffic flow gradually decreases, and the average fuel consumption and emissions of four pollutants corresponding to all control strategies gradually increase. As the penetration rate of CAVs increases, the fuel consumption and pollutant emissions corresponding to the BS-BS strategy gradually increase, while the other nine strategies are the opposite. In low-density traffic environments, applying different control strategies has little effect on reducing the average fuel consumption and pollutant emissions, which are relatively low. However, when the traffic environment enters the medium to high-density stage, under a penetration rate of 20%, the fuel consumption under the CTG-CTG strategy is relatively low. As the penetration rate of CAVs increases, the traffic flow tends to stabilize. Except for the BS-BS strategy, the average speed of traffic flow using other strategies gradually increases, and the fuel consumption decreases. When the penetration rate of CAVs is greater than or equal to 80%, the fuel consumption corresponding to the VTG1-CS, VTG2-CS, and CTG-CS strategies is lower, the BS-BS strategy is the highest, and the CTG-CTG strategy is the second highest.

(2) CAVs have strong stability, but their addition does not continuously improve the stability of mixed traffic flow or reduce fuel consumption. Under conditions of low penetration rate and medium to high density, it can cause significant speed disturbances within the traffic flow, exacerbating instability and increasing fuel consumption. At this point, it is recommended to use CTG, VTG, VTG-CTG, and CTG-CS strategies.

(3) The unstable mixed traffic flow has a relatively weak impact on increasing pollutant emissions. In terms of $CO_2$ emissions, under a low penetration rate, there is little difference among the nine strategies except for the BS-BS strategy. However, when the penetration rate of CAV is greater than 40%, the $CO_2$ emissions corresponding to the three combination strategies, VTG1-CS, VTG2-CS, and CTG-CS, are lower than those of other strategies, with the CTG-CTG strategy being the highest. In terms of NOx and VOC emissions, the pollutant emissions corresponding to VTG1-CS, VTG2-CS, and CTG-CS strategies are generally lower than those of other strategies, with BS-BS strategy being the worst and CTG-CTG strategy second. Regarding PM emissions, when the CAV penetration rate is greater than or equal to 60%, the average PM emissions of traffic flow using VTG1-CS, VTG2-CS, and CTG-CS strategies are lower. Under low penetration rates, the VTG1-CS and VTG2-CS strategies result in more PM emissions from mixed traffic flow, while the CTG-CS strategy outperforms these two strategies. The BS-CS strategy results in higher PM emissions than the VTG1-CS and VTG2-CS strategies in medium-density environments, but the difference between the BS-CS strategy and the VTG1-VTG1 strategy is minimal in high-density environments. The other four strategies are not significantly different from the VTG1-VTG1 strategy.

However, this work still has limitations. Based on these limitations, future research can be conducted.

(1) Developing a car-following model that considers time delay. CAVs can use V2V communication technology and onboard sensors to obtain operational information from other vehicles. However, both technologies have a specific time delay in real. Due to this delay being almost negligible compared to the reaction time of human drivers, it was not considered in this paper. However, establishing a car-following model that considers time delay is more practical and helpful for implementing CAVs control strategies.

(2) Investigating the heterogeneous traffic flow mixed in by CAVs with different levels of autonomous driving. The existing authoritative standards classify vehicle autonomous driving levels into six levels, L0 to L5. Due to the large number of control strategies considered in this paper, to simplify the complexity of the problem, CAVs are set to the same autonomous driving level. However, in subsequent expansion research, car-following models can be equipped for CAVs with different autonomous levels.

**Acknowledgements**

The paper received research funding support from the National Natural Science Foundation of China (72471200), the Sichuan Science and Technology Program (2024NSFSC0179), the Fundamental Research Funds for the Central Universities (2682023ZTPY034), and the Chengdu Soft Science Research Project (2023-RK00-00029-ZF).